\documentclass[tightenlines,twoside,secnumarabic,
               onecolumn,floatfix,nofootinbib,preprint,11pt]{revtex4-1}
\usepackage{graphicx}
\usepackage{dcolumn}
\usepackage{float}
\usepackage{amsmath}
\usepackage{amsfonts}
\usepackage{multirow}
\usepackage{color}

\begin{document}
\def\Lt{\overline{L}}
\def\nn{{++}}
\def\an{{+-}}
\def\bn{{+\pm}}
\def\sl{{\rm sl}}
\def\ax{{\rm ax}}
\def\vect{{\rm v}}
\def\ib{{\; \bar\imath}}
\def\Atree{A^{\rm tree}}
\def\half{\textstyle \frac{1}{2}}
\def\nf{n_{\mskip-2mu f}}
\def\q{{q}}
\def\Q{{Q}}
\def\qb{{{\bar q}}}
\def\Qb{{{\bar Q}}}
\def\sl{{\rm sl}}
\def\ax{{\rm ax}}
\def\A#1{{\cal A}_{#1}}
\def\vect{{\rm v}}
\def\cM{\mathcal{M}}
\def\prop#1{{\cal P}_{#1}}
\def\Af{v_A^f}
\def\Vf{v_V^f}
\def\fudgeBDK{-}
\def\mfudgeBDK{\relax}
\def\tree{{\rm tree}}
\def\sutqm{(s_{13}+s_{14})}
\def\sdtqm{(s_{23}+s_{24})}
\def\sucsm{(s_{15}+s_{16})}
\def\sdcsm{(s_{25}+s_{26})}
\def\sutq{{s}_{134}}
\def\sdtq{{s}_{234}}
\def\sucs{{s}_{156}}
\def\sdcs{{s}_{256}}
\def\sud{{s}_{12}}
\def\stq{{s}_{34}}
\def\scs{{s}_{56}}
\def\sut{{s}_{13}}
\def\suq{{s}_{14}}
\def\sdt{{s}_{23}}
\def\sdq{{s}_{24}}
\def\suc{{s}_{15}}
\def\sus{{s}_{16}}
\def\sdc{{s}_{25}}
\def\sds{{s}_{26}}
\def\duxtqxd{1}
\def\ddxuxtq{2}
\def\duxdxtq{3}
\def\cuxd{1}
\def\cudxtq{2}
\def\cuxtq{3}
\def\cdxtq{4}
\def\cuxcs{5}
\def\cdxcs{6}
\def\cD{\cal D}
\def\cP{\cal P}
\def\cg{c_\Gamma}
\def\gW{g_W}
\def\rG{r_\Gamma}
\def\jb{i_{b}}
\def\jt{i_{t}}
\def\tb{\bar{t}}
\def\eb{\bar{e}}
\def\mub{\bar{\mu}}
\def\qb{\bar{q}}
\def\spaa#1.#2.#3{\langle\mskip-1mu{#1}|#2|{#3}\mskip-1mu\rangle}
\def\spbb#1.#2.#3{[\mskip-1mu{#1}|#2|{#3}\mskip-1mu]}
\def\spa#1.#2{\left\langle#1\,#2\right\rangle}
\def\spb#1.#2{\left[#1\,#2\right]}
\def\spab#1.#2.#3{\left\langle#1|#2|#3\right]}
\def\spba#1.#2.#3{\left[#1|#2|#3\right\rangle}
\def\spbab#1.#2.#3.#4{[\mskip-1mu{#1}
                  | #2  #3 | {#4}\mskip-1mu]}
\def\spaba#1.#2.#3.#4{\langle\mskip-1mu{#1}
                  | #2  #3 | {#4}\mskip-1mu\rangle}
\def\Wzb{\bar{W}_0}
\def\Pzb{\bar{P}_0}
\def\Pthreeb{\bar{P}_3}
\def\P3b{\bar{P}_3}
\def\Ypb{\bar{Y}_p}
\def\Ypz{{Y}_p(z)}
\def\Ywb{\bar{Y}_w}
\def\Ppb{\bar{P}_+}
\def\Pmb{\bar{P}_-}
\def\Wpb{\bar{W}_+}
\def\Wmb{\bar{W}_-}
\def\cO{{\cal O}}
\def\li{{\rm Li_2}}
\def\g0{\gamma_0}
\def\gp{\gamma^{+}}
\def\gm{\gamma^{-}}
\def\lp{\gamma^{+}}
\def\lm{\gamma^{-}}
\def\xp{x_{+}}
\def\xm{x_{-}}
\def\bentarrow{\:\raisebox{1.3ex}{\rlap{$\vert$}}\!\rightarrow}                 
\def\dkp#1#2#3#4{
\begin{array}{r c l}
#1 & \rightarrow & #2#3 \\
 & & \phantom{\; #2}\bentarrow #4
\end{array}}                                                                    
\def\bothdk#1#2#3#4#5{
\begin{array}{r c l}                                                            
#1 & \rightarrow & #2#3 \\
 & & \:\raisebox{1.3ex}{\rlap{$\vert$}}\raisebox{-0.5ex}{$\vert$} 
\phantom{#2}\!\bentarrow #4 \\
 & & \bentarrow #5                                                              
\end{array}                                                                     
}                                                                               
\newcommand{\kirill}{\colour{red}}
\newcommand{\comment}[1]{{\bf [#1]}}
\newcommand{\beq}{\begin{equation}}
\newcommand{\eeq}{\end{equation}}
\newcommand{\beqn}{\begin{eqnarray}}
\newcommand{\eeqn}{\end{eqnarray}}
\newcommand{\bi}[1]{\bibitem{#1}}
\newcommand{\fr}[2]{\frac{#1}{#2}}
\newcommand{\non}{\nonumber}
\newcommand{\Et}{E_t}
\newcommand{\Pt}{P_t}
\newcommand{\pt}{p_t}
\newcommand{\pb}{p_b}
\newcommand{\pw}{p_W}
\newcommand{\pg}{p_g}
\newcommand\tpW        {{\tilde p}_W}
\newcommand\tpb        {\tilde p_b}
\newcommand{\ar}{\mbox{$\rightarrow$}}
\def\ra{\rightarrow}

\newcommand{\slsh}{\rlap{$\;\!\!\not$}}     
\def\amuh{a_\mu^{{\mathrm had}}}
\def\vec#1{{\mbox{\boldmath$#1$}}}
\def\ket#1{\vert #1 \rangle}
\def\bra#1{\langle #1 \vert}
\newcommand{\as}{\alpha_S}
\newcommand{\p}{\mbox{$\vec{p}$}}
\newcommand{\pp}{\mbox{$\vec{p}'$}}
\newcommand{\rp}{\mbox{$\vec{r}'$}}
\newcommand{\kp}{\mbox{$\vec{k}'$}}
\newcommand{\e}{\mbox{$\vec{e}$}}
\newcommand{\s}{\mbox{$\vec{s}$}}
\newcommand{\Li}{{\rm Li}}
\newcommand{\lsim}{\mbox{\raisebox{-0.3ex}{%
\footnotesize $\:\stackrel{<}{\sim}\:$}} }
\newcommand{\gsim}{\mbox{\raisebox{-0.3ex}{%
\footnotesize $\:\stackrel{>}{\sim}\:$}} }
\newcommand{\lb}{\left (}
\newcommand{\rb}{\right )}
\newcommand{\ep}{\epsilon}
\newcommand{\vep}{\epsilon}
\newcommand{\dd}{{\rm d}}
\newcommand{\om}{\omega}

\newcommand{\sS}{\mbox{$\vec{\sigma}\vec{\sigma}'$}}
\newcommand{\si}{\mbox{$\vec{\sigma}$}}
\newcommand{\vgamma}{\mbox{$\vec{\gamma}$}}
\newcommand{\vxi}{\mbox{$\vec{\xi}$}}

\newcommand{\pop}[1]{\mbox{$\Lambda_+(#1)$}}
\newcommand{\nep}[1]{\mbox{$\Lambda_-(#1)$}}
\newcommand{\Dafne}{DA$\Phi$NE}
\newcommand{\mar}{\marginpar{***}}
\def\spab#1.#2.#3{\langle\mskip-1mu{#1}
                  | #2 | {#3}\mskip-1mu]}
\def\spba#1.#2.#3{[\mskip-1mu{#1}
                  | #2 | {#3}\mskip-1mu\rangle}
\def\spa#1.#2{\langle#1\,#2\rangle}
\def\spb#1.#2{[#1\,#2]}

\def\dk#1#2#3{
\begin{array}{r c l}
#1 & \rightarrow & #2 \\
 & & \bentarrow #3
\end{array}
}

\title{Top-quark loop corrections in $Z+$jet and $Z+2$~jet production}

\author{John M. Campbell}
\email{johnmc@fnal.gov}
\affiliation{Fermilab, Batavia, IL 60510, USA}
\author{R. Keith Ellis}
\email{keith.ellis@durham.ac.uk}
\affiliation{Institute for Particle Physics Phenomenology,
Department of Physics, Durham University, Durham DH1 3LE, United Kingdom}
\preprint{FERMILAB-PUB-16-429-T,\, IPPP/16/88}

\begin{abstract}
The sophistication of current predictions for $Z+$jet production at hadron colliders
necessitates a re-evaluation of any approximations inherent in the theoretical calculations.
In this paper we address one such issue, the inclusion of mass effects in top-quark loops.
We ameliorate an existing calculation of $Z+1$~jet and $Z+2$~jet production by presenting exact
analytic formulae for amplitudes containing top-quark loops that enter at next-to-leading order
in QCD.  Although approximations based on
an expansion in powers of $1/m_t^2$ can lead to poor high-energy behavior, an exact treatment of
top-quark loops demonstrates that their effect is small and has limited phenomenological interest.
\end{abstract}
\maketitle

\section{Introduction}

The production of a $Z$-boson in association with jets is of considerable importance as a tool for understanding the Standard Model (SM).
The $Z+$jet process has been proposed as a probe of both parton distribution functions and the high-energy running of the
strong coupling, $\alpha_s$.   The production of more than one jet is especially important in the environment of the LHC, where
typical jet reconstruction algorithms routinely result in multiple jets.  This means that $Z$+jets processes represent significant
backgrounds in many searches for New Physics, notably when the $Z$-boson decays to neutrinos so that it is a source of large
missing transverse energy.  Therefore they must be predicted precisely within the SM.

In order to obtain the level of theoretical precision required to match the small experimental uncertainties~\cite{Aad:2013ysa,Khachatryan:2014zya},
it is imperative to perform perturbative calculations of $Z+$jet processes beyond the leading order.   The dominant source of
corrections arises from QCD, with next-to-leading order (NLO) calculations available for processes involving up to four
jets~\cite{Giele:1993dj,Campbell:2002tg,Campbell:2003hd,Berger:2010vm,Ita:2011wn}.  The expected experimental precision of
measurements of the $Z+1$~jet state has motivated the calculation of this process to the next order
(NNLO)~\cite{Ridder:2015dxa,Boughezal:2015ded,Ridder:2016nkl,Boughezal:2016yfp},
so that experimental and theoretical uncertainties in this case are commensurate over a relatively large kinematic range.
At this level of theoretical accuracy it is also necessary to have control over corrections arising from the electroweak
sector.  These effects are known for up to two jets in the final state~\cite{Kuhn:2005az,Denner:2011vu,Kallweit:2015dum}.

With these results in hand it is important to revisit assumptions and approximations inherent in the calculations performed
so far.  One such approximation relates to the inclusion of the effect of the top quark in one-loop virtual corrections to these
processes.    Since the mass of the top quark introduces a new scale into the problem, including its effect results in a 
significantly more complex calculation than the usual case in which all quarks are considered massless.  In their classic
1997 paper~\cite{Bern:1997sc}, Bern, Dixon and Kosower (BDK) gave results for such contributions to the $Z+1$ and $Z+2$~jet
processes by performing a large mass expansion in the top-quark mass. Although this approximation was appropriate in the last century,
and in particular for $e^+ e^-$
annihilation at LEP energies, it may no longer be appropriate at the LHC and higher energy machines where scales above
the top quark mass are probed.

In this paper we shall compute a class of one-loop corrections to $Z+1$ and $Z+2$~jet processes,
specifically considering the effects of fermion loops in which the full dependence on the top-quark
mass is retained. The one-loop results for these processes can be obtained with a number of numerical programs,
but it is nevertheless useful to have analytic formulae for these processes, because of the improvement
in evaluation speed and numerical stability that an analytic formula can provide.  The amplitudes that we have
computed may be useful for other crossed, or related, processes and are provided in the appendix.  The
phenomenological impact of these calculations is assessed for the $Z+1$~jet case in Section~\ref{sec:z1jresults}
and for the $Z+2$~jet process in Section~\ref{sec:z2jresults}.

\section{Top-loop effects in $Z+1$~jet production}
\label{sec:z1jresults}

\begin{figure}
\includegraphics[angle=270,width=0.8\textwidth]{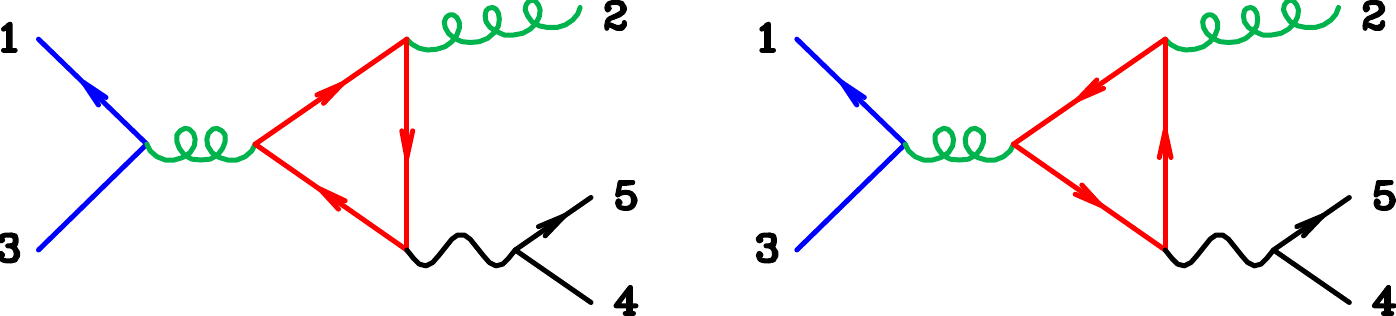}
\caption{Examples of fermion loop diagrams contributing to $Z+1$~jet production.  The only non-zero contribution
enters through the axial coupling of the $Z$-boson to third-generation quarks.}
\label{floop5}
\end{figure}

In the case of $Z+1$~jet production, top-quark loop contributions only
enter through diagrams such as the ones shown in Fig.~\ref{floop5}.
Furry's theorem means that diagrams containing a vector coupling of
the $Z$-boson to the quark loop vanish, so that only the axial
coupling contributes.  In fact, since we consider all quarks other than 
the top quark to be massless, 
due to the opposite weak isospin of up- and
down-type quarks, the only contribution from these diagrams comes from the
third generation.
In the original BDK treatment of these diagrams~\cite{Bern:1997sc}, these contributions are
computed in the limit that $m_t \to \infty$, with the leading term in a $1/m_t^2$ expansion
retained.  We have recomputed these contributions retaining the full top-quark mass dependence;  the analytic form of
the amplitudes representing this contribution is given in Appendix~\ref{sec:fivepoint}.

\begin{figure}
\includegraphics[angle=0,width=0.49\textwidth]{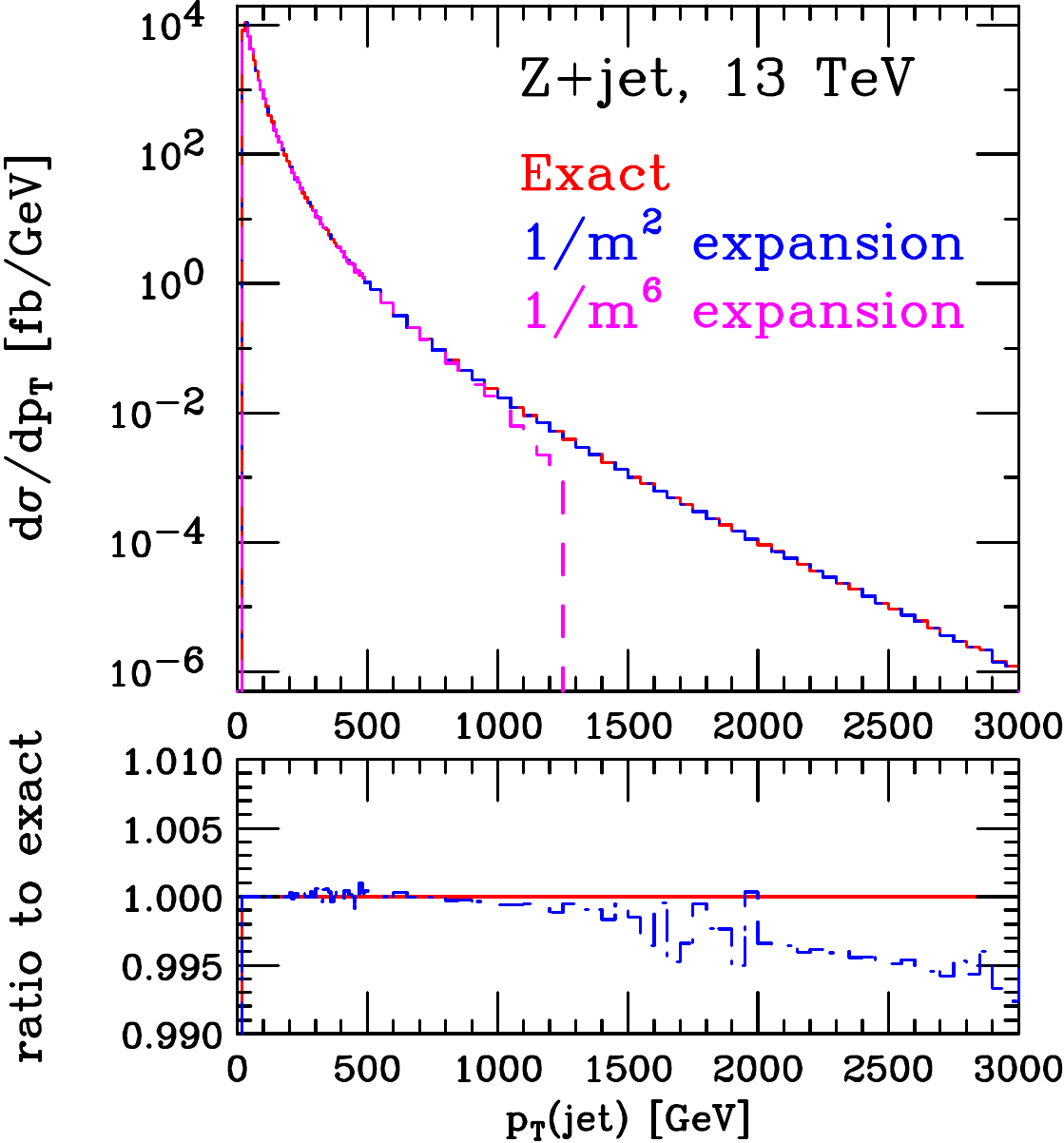}
\includegraphics[angle=0,width=0.49\textwidth]{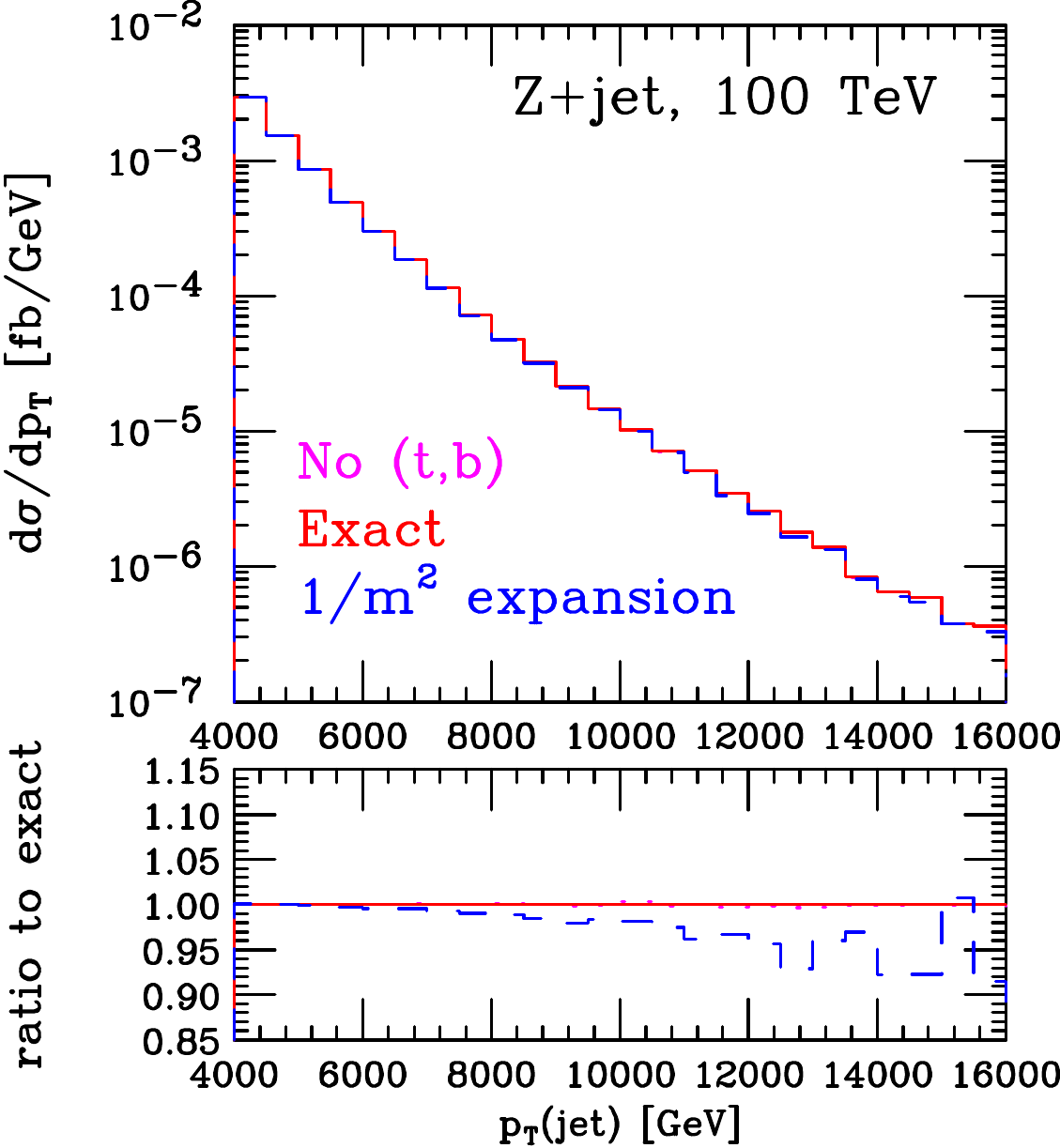}
\caption{The jet $p_T$ spectrum for $Z+$jet production at NLO, computed for the 13 TeV LHC (left) and
a 100 TeV $pp$ collider (right).  The calculation uses a scale $\mu_r = \mu_f = H_T/2$ and no cuts are applied apart from
$p_T(\mbox{jet})>25$~GeV.  The red (solid) histogram corresponds to the exact result while the blue (dot-dash)
and magenta (dash) histograms represent the large-mass expansion up to $1/m_t^2$ and $1/m_t^6$ respectively,
as detailed in the text.
\label{z1j}}
\end{figure}

This expansion can be extended to include higher-order terms but in the high-energy
regime this can lead to problems since the expansion is properly
of the form $s/m_t^2$, where $s$ becomes large.  This is illustrated in Fig.~\ref{z1j} (left).  The leading
term in the expansion (as presented in BDK) agrees very well with the exact result over the range shown.
Including further terms in the $1/m_t^2$ expansion spoils this agreement.  Although
the exact treatment and the $1/m_t^6$ approximation agree up to jet transverse momenta around 1 TeV, beyond that
the approximation is no longer under control and results in a wildly different prediction for
the spectrum.  The lower panel shows the ratio of the approximation with the leading term to the exact result.
The two differ by around $0.7\%$ for a jet with $3$~TeV transverse momentum.  Since the number of events in this region
is negligible this is not a significant difference.
We conclude that, although the exact result should be preferred, there is no observable impact on the phenomenology
of this process when using only the leading term in the $1/m_t^2$ expansion.

At a 100 TeV collider the differences are more significant, as shown in Fig.~\ref{z1j} (right).
Even for 10 TeV jets, which would be abundant
at such a collider, the effect of the approximate top-quark loop is a few percent.  Since this is at the
same level as the NNLO corrections, it is important that the exact result be available and taken into account.

\section{Top-loop effects in $Z+2$~jet production}
\label{sec:z2jresults}

The $Z+2$~jet process is sensitive to a much wider range of virtual corrections that involve a closed loop of
top quarks.  This is partly due to the fact that the process is represented by two separate parton-level
reactions (and all appropriate crossings):
\beqn
0 &\to & q(-p_1)+\bar{q}(-p_2)+g(p_3)+g(p_4)+e^{+}(p_5)+e^{-}(p_6)\, , \\
0 &\to & q(-p_1)+\Qb(-p_2)+Q(p_3)+\bar{q}(p_4)+e^{+}(p_5)+e^{-}(p_6)\, .
\eeqn
We will refer to these by the abbreviated forms, $q\bar q gg Z$ and $q\bar q Q\bar Q Z$ processes.  In the
original BDK paper, all of the top-quark loop contributions have been included using the $1/m_t^2$ expansion.

In this work we have computed all of the corrections retaining the full dependence on the top-quark mass.  The
addition of the mass complicates the analytic form of the amplitudes but we have still obtained relatively
compact expressions.  This is achieved through the use of analytic unitarity methods for computing one-loop
box and triangle coefficients~\cite{Britto:2004nc,Britto:2006sj,Badger:2008cm} and by recycling BDK results for
the massless case whenever possible.  Full details of our calculation, including explicit expressions for all
amplitudes, are presented in Appendices~\ref{sec:sixpointqqQQ} and~\ref{sec:sixpointqqgg}.

\begin{figure}
\begin{center}
\includegraphics[angle=270,width=0.7\textwidth]{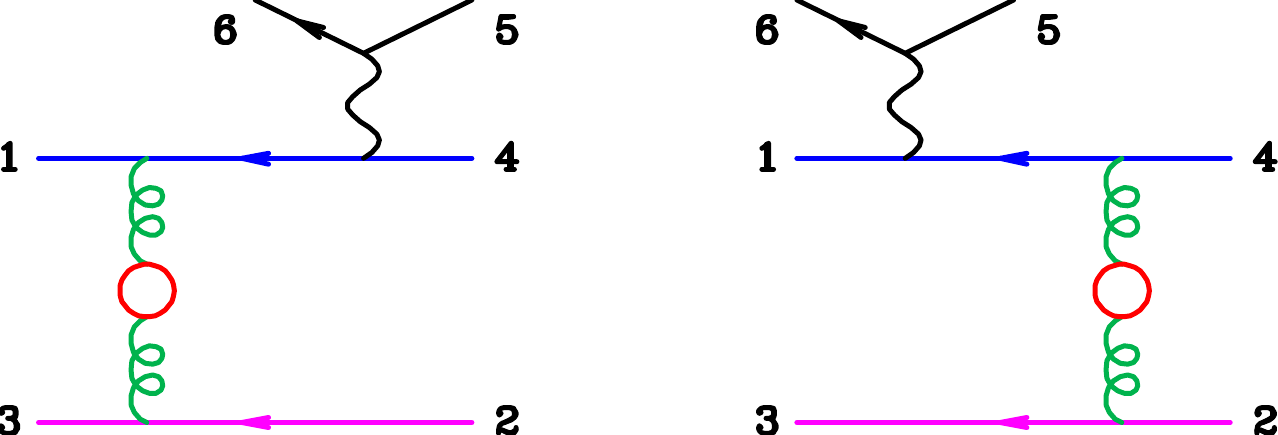}
\caption{Vacuum polarization diagrams contributing to the $q\bar q Q \bar Q Z$ process,
where the $Z$-boson couples to an external line of light quarks.}
\label{floopqqQQa}
\end{center}
\end{figure}
\begin{figure}
\includegraphics[angle=270,width=0.9\textwidth]{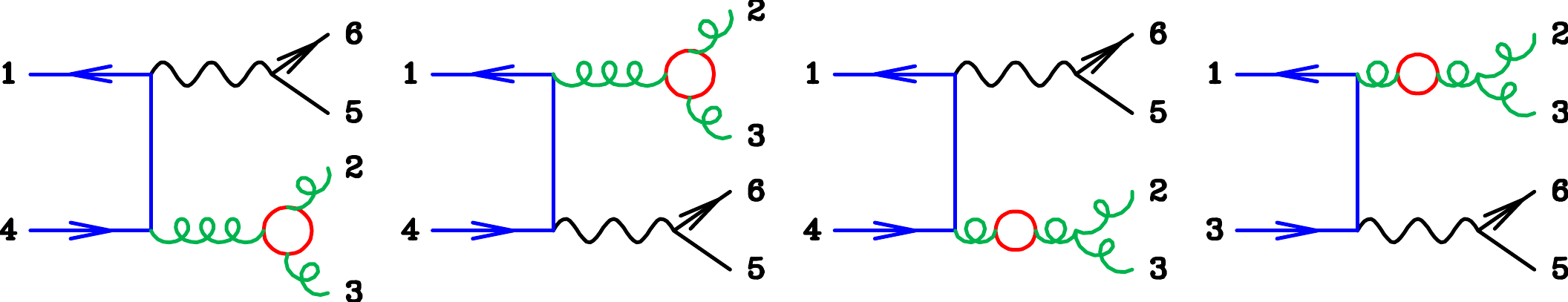}
\caption{Examples of fermion loop (top quark) diagrams in which the $Z$-boson couples to an external line of light quarks
in the $q\bar q gg Z$ process.
}
\label{floopa}
\end{figure}
\begin{figure}
\includegraphics[angle=270,width=0.9\textwidth]{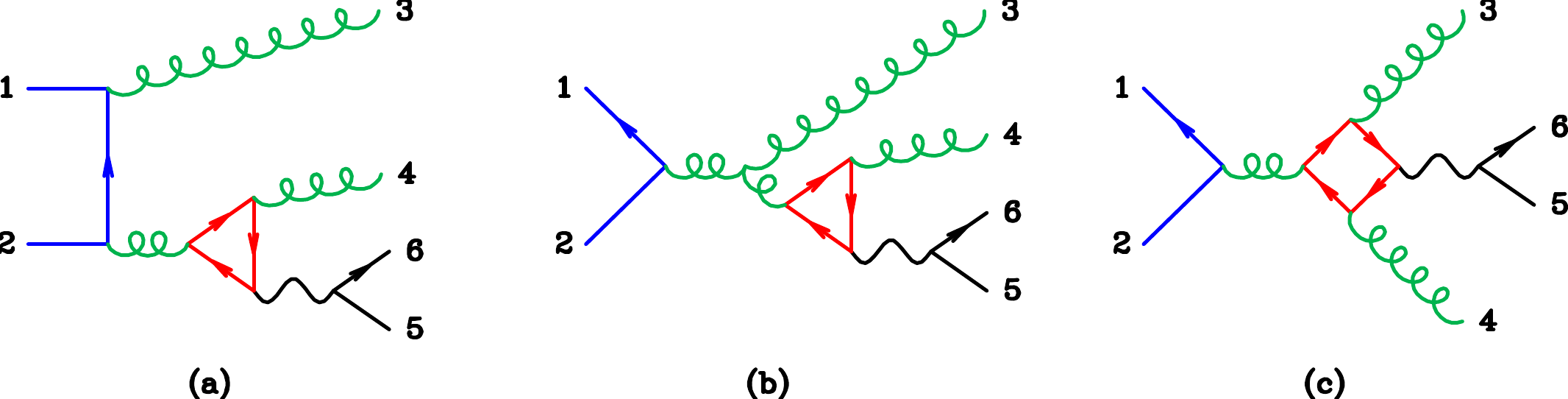}
\caption{Examples of fermion loop diagrams contributing when the $Z$ couples 
to a heavy quark line through either a vector or an axial coupling.
With a vector coupling the triangle diagrams vanish, and hence only the box diagrams contribute.}
\label{floopb}
\end{figure}
\begin{figure}
\includegraphics[angle=270,width=0.7\textwidth]{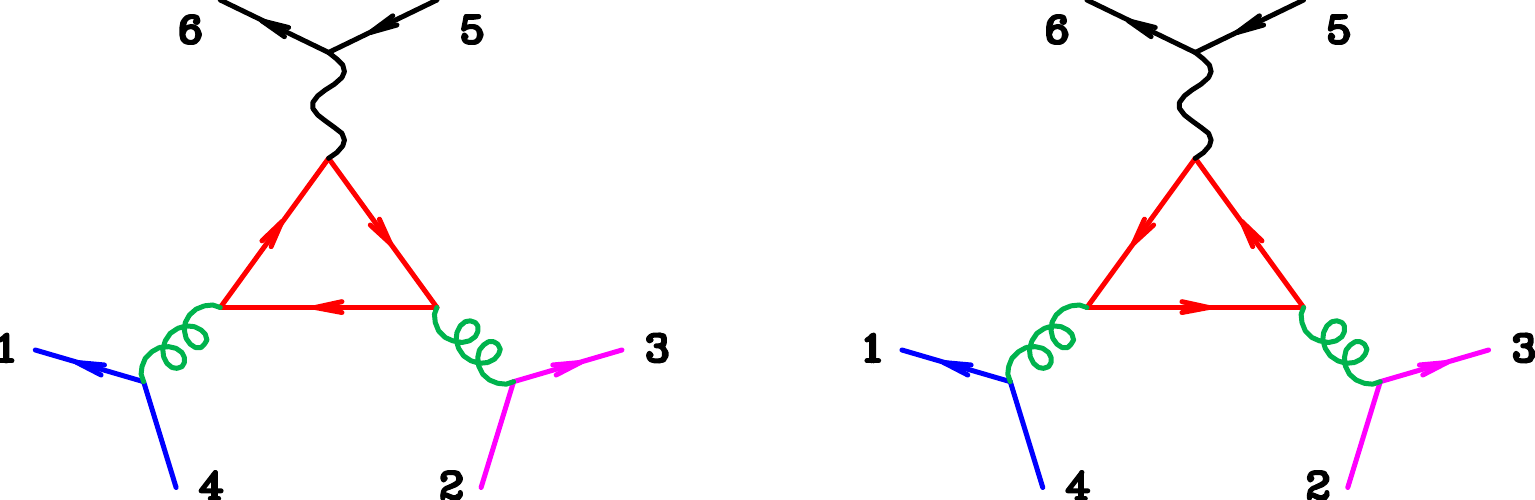}
\caption{Quark loop diagrams involving the axial coupling of the $Z$-boson in the $q \bar q Q \bar Q Z$ process.}
\label{floopqqQQb}
\end{figure}

The top-quark loop contributions can be categorized according to the manner in which the $Z$-boson couples to the partons:
\begin{enumerate}
\item Contributions where the $Z$ boson couples to the light quark line.  These correspond to
vacuum polarization contributions to the $q\bar q Q\bar Q Z$ process shown in Fig.~\ref{floopqqQQa}
and to the loop corrections to the $q\bar q gg Z$ process depicted in Fig.~\ref{floopa}.  These
amplitudes are described in detail in Appendices~\ref{sec:sixpointqqQQ}.\ref{sec:qqQQvacpol} and~\ref{sec:sixpointqqgg}.\ref{sec:qqggcouplelight}.
\item A vector coupling of the $Z$-boson to a closed loop of top quarks, occurring in diagrams such
as the one shown in Fig.~\ref{floopb}(c).  These are only present in the $q\bar q gg Z$ process and are
described in Appendix~\ref{sec:sixpointqqgg}.\ref{sec:qqggvector}.
\item An axial coupling of the $Z$-boson to a closed loop of quarks, as shown in Figs.~\ref{floopb} and~\ref{floopqqQQb}
for the $q\bar q gg Z$ and $q\bar q Q\bar Q Z$ processes, respectively.  This contribution vanishes for all but
the third generation of quarks, whose effect is captured here.  For the $q\bar q Q\bar Q Z$ process these
corrections are discussed in Appendix~\ref{sec:sixpointqqQQ}.\ref{sec:qqQQaxial} while the corresponding contributions to
the  $q\bar q gg Z$ process are detailed in Appendices~\ref{sec:sixpointqqgg}.\ref{sec:qqggaxialsl} and~\ref{sec:sixpointqqgg}.\ref{sec:qqggaxial}.
\end{enumerate}

We will now  examine the effect of each of these contributions separately, both in the $1/m_t^2$ approximation used
in the BDK form of the amplitudes and with the improved treatment provided by the exact expressions presented here.
Our calculation is performed by incorporating our newly-calculated amplitudes in the Monte Carlo program
MCFM~\cite{Campbell:1999ah,Campbell:2011bn,Campbell:2015qma}, which already includes a complete calculation
of $Z+2$~jet production at NLO that makes use of the BDK loop amplitudes.
The expressions for the amplitudes with the exact top-mass dependence are written in terms of the scalar integrals
described in Appendix~\ref{app:scalarintegrals}, that are evaluated numerically using
the {\tt ff}~\cite{vanOldenborgh:1989wn,vanOldenborgh:1990yc} and {\tt QCDLoop}~\cite{Ellis:2007qk,Carrazza:2016gav}
libraries.

For all of the results in this section we will consider the production of an on-shell $Z$-boson that decays
to an electron-positron pair, with no cuts applied to the leptons.  This is a choice made for the presentation 
of our results, and not an intrinsic limitation of MCFM. 
We use the {\tt CT14.NN} pdf set~\cite{Dulat:2015mca}
and choose to set both renormalization and factorization scales to $H_T/2$, where $H_T$ is the scalar sum of the
transverse momenta of all leptons and partons.

\subsection{Results: $100$~TeV collider}
\begin{figure}
\includegraphics[angle=0,width=0.7\textwidth]{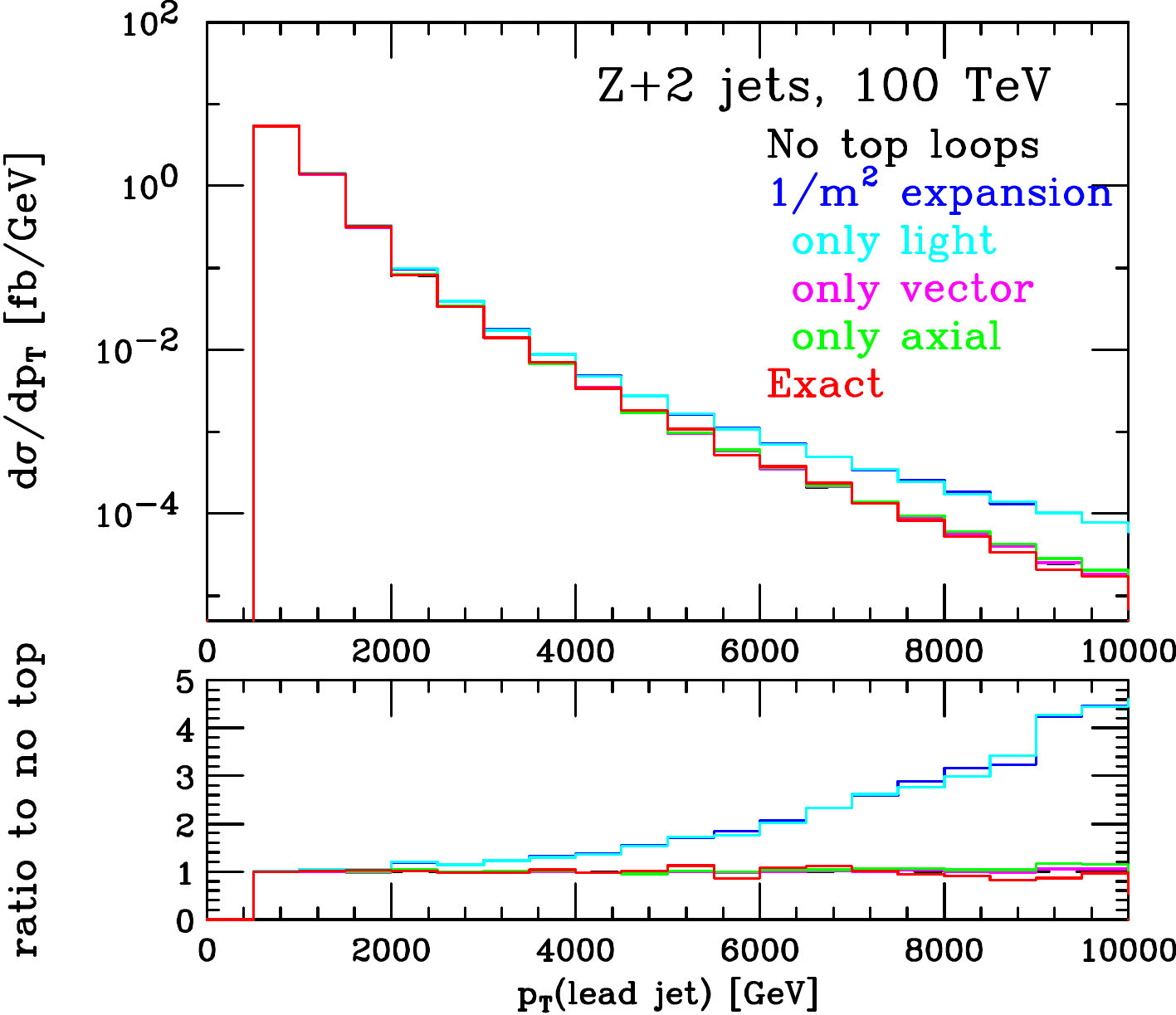}
\caption{Upper panel: the distribution of the transverse momentum of the leading jet in $Z+2$~jet events at $100$~TeV.
Predictions are shown with no top-quark loops included, using the $1/m_t^2$ approximation and with the exact result
(including all contributions).  The exact, only vector and only axial histograms are almost indistinguishable from
the result with no top-quark loops.    Lower panel:  the ratio of the predictions of the approximate treatment to the one in
which no top-quark loops are included.}
\label{fig:z2j100}
\end{figure}

Since we expect the problems associated with the $1/m_t^2$ expansion used in the original BDK expressions to be
exacerbated at high energies, we first present results for a putative $100$~TeV proton-proton collider.  We define
jets using the anti-$k_T$ clustering algorithm with a jet separation $R=0.5$ and demand that they satisfy,
\begin{equation}
p_T(\mbox{jet}) > 500~\mbox{GeV} \,, \qquad
y(\mbox{jet}) < 4 \,.
\end{equation}
A comparison of the NLO predictions for the lead jet transverse momentum, with various levels of sophistication,
is shown in Fig.~\ref{fig:z2j100}.  The approximation of the contributions with the $Z$-boson coupled to a
top-quark loop through axial and vector couplings lead to relatively small deviations in this range.  In contrast,
approximating the contributions that involve the $Z$-boson coupling to light quarks in the same way leads to
substantial errors at jet transverse momenta of about $3$~TeV and higher.  The NLO rate is over-estimated by a
factor of four for a $10$~TeV jet.  Using the exact result for the top-quark loops yields a prediction that is
essentially unchanged from the one in which they are not included at all.

\subsection{Results: LHC at $\sqrt s=14$~TeV}
\begin{figure}
\includegraphics[angle=0,width=0.7\textwidth]{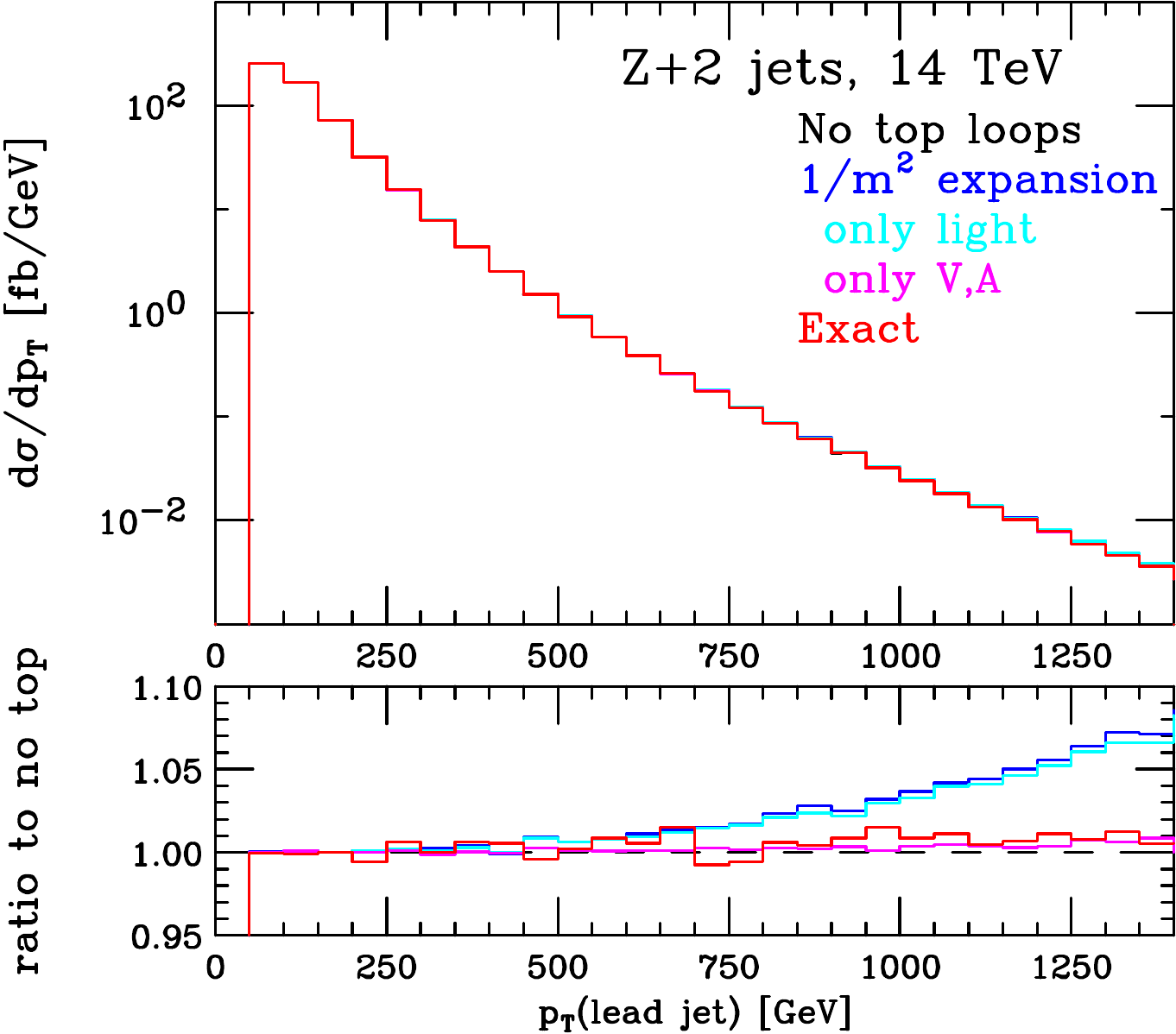}
\caption{Upper panel: the distribution of the transverse momentum of the leading jet in $Z+2$~jet events at $14$~TeV.
Predictions are shown with no top-quark loops included, using the $1/m_t^2$ approximation and with the exact result
(including all contributions).  The exact, and only vector and axial, histograms are barely distinguishable from
the result with no top-quark loops.    Lower panel:  the ratio of the predictions of the approximate treatment to the one in
which no top-quark loops are included.}
\label{fig:z2j14}
\end{figure}

We now turn to results of more immediate interest, namely predictions for the LHC operating at $\sqrt s=14$~TeV.
We adjust the jet cuts accordingly and now demand,
\begin{equation}
p_T(\mbox{jet}) > 50~\mbox{GeV} \,, \qquad
y(\mbox{jet}) < 2.5 \,.
\end{equation}
A comparison of our calculations under these cuts is shown in Fig.~\ref{fig:z2j14}.  Note that, in comparison to the
previous figure, the lower panel has a much smaller scale since we consider transverse momenta for the jet that are
much lower.  In addition, having observed that the effect of the diagrams in which the $Z$-boson couples to a
top-quark loop is small, in this case we simply show the sum of the contributions from the vector and
axial couplings of the $Z$-boson to top quarks. As expected, at the energies that are accessible at the LHC the 
error made when using the $1/m_t^2$
approximation is much less severe.  Even at a jet transverse momentum of $1$~TeV it only results in a 4\% deviation
from the result with no top-loops included.  The effect of the approximation on the cross-section for both jets
above $50$~GeV is an enhancement of a mere $0.05$\%.

\section{Conclusions}

In this paper we have reviewed the importance of top-quark loops in NLO corrections to
$Z+1$~jet and $Z+2$~jet production.  To do so, we have computed the effect of these loops
with an exact treatment of the top-quark mass and given analytic forms for all the relevant
amplitudes.  We find that the effect of these loops is
very small and not important for phenomenology at the LHC.  For a putative $100$~TeV
proton-proton collider the effects are more significant and, for the $Z+1$~jet case, lead to a few percent
change in the prediction for jets with transverse momentum of $10$~TeV.  Attempting to include the effect of these loops by
using an expansion in powers of $1/m_t^2$ leads to the theoretical prediction being over-estimated
due to poor high-energy behaviour.  While this may be at a level that is tolerable at the LHC,
it can lead to results at $100$~TeV that are incorrect by factors of two or more.

\section*{Acknowledgements}
RKE would like to thank the Fermilab theory group for hospitality during 
the preparation of this paper. 
The research of JMC is supported by the US DOE under contract DE-AC02-07CH11359.

\bibliography{Z1jet_mt}

\appendix

\section{Five point amplitude $A(1_q,2_g,3_{\qb},4_{\eb},5_e)$.}
\label{sec:fivepoint}

In this appendix we consider the five-point amplitudes that enter the calculation of
$Z+1$~jet production.  Specifically, we consider the process:
\beq
0 \to  q(-p_1)+g(p_2)+\bar{q}(-p_3)+e^{+}(p_4)+e^{-}(p_5)\, .
\eeq

\subsection{Tree graphs}
We write the tree-level amplitude as,
\beq
{\cal A}_5^{\rm tree}=2 e^2 g \big(-Q^q+v^e_{L,R} v^q_{L,R} \prop{Z}(s_{45})\big) \; \big(T^{a_2}\big)_{i_1}^{~\ib_3} \; 
A_5^{\rm tree}(1_q,2_g,3_{\qb}) \,,
\eeq
where we have omitted the labels of the electron-positron pair, ($5$ and $4$ respectively).
We further define
\beq
s_{ij}=(p_i+p_j)^2,\;\;\;
s_{ijk}=(p_i+p_j+p_k)^2\, .
\eeq
$e$ is the QED coupling, $g$ the QCD coupling, $Q^\q$ is the charge of
quark $q$ in units of $e$, (the positron charge), 
and the ratio of $Z$ and photon propagators
is given by
\begin{equation}
\prop{Z}(s) = {s \over s - M_Z^2 + i \,\Gamma_Z \, M_Z}\,,
\end{equation}
where $M_Z$ and $\Gamma_Z$ are the mass and width of the $Z$.
The definition of the $Z/\gamma^*$ couplings is given in Table~\ref{Feynmanrules}.
Colour matrices are normalized such that 
\beq
{\rm Tr}\, T^{a_1} T^{a_2} = \delta^{a_1 \, a_2} \, .
\eeq
\renewcommand{\baselinestretch}{1.2}
\begin{table}
\begin{tabular}{|c|c|c|}
\hline
Boson & Feynman rule & Coupling\\
\hline
$\gamma$  &  $ -i e Q^f \gamma^\mu $  &   \\
\multirow{ 2}{*}{$Z$} &  
$ -i e \gamma^\mu (\Vf -\Af \gamma_5) $
& $\Vf = (\tau_3^f -2 Q^f \sin^2 \theta_W)/\sin 2 \theta_W$,
\; $ \Af = \tau_3^f/\sin 2 \theta_W, \;\;\tau_3^f =\pm \frac{1}{2}$ \\
&$  -i e \gamma^\mu (v_L^f \gamma_L + v_R^f \gamma_R)$ & $v_L^f =(2 \tau_3^f - 2 Q^f\sin^2 \theta_W)/\sin 2 \theta_W  $, \;
$v_R^f  = -2 Q^f \sin^2 \theta_W /\sin 2 \theta_W $  \\
\hline
\end{tabular}
\caption{Feynman rules and couplings of a photon and a $Z$ to a fermion-antifermion pair. 
For massless fermions it is convenient to
use the left- and right-handed couplings, rather than the vector and axial couplings, so both are shown.
$Q^f$ is the charge of the fermion in units of the positron electric charge.}
\label{Feynmanrules}
\end{table}
\renewcommand{\baselinestretch}{1}
For the tree amplitude 
$A_5(1_q^+ , 2_g^+ , 3_{\qb}^- , 4_{\eb}^- , 5_e^+)$, 
the result is,
\beq
A_5^\tree(1_q^+,2_g^+,3_{\qb}^-) = -i \frac{{\spa3.4}^2}{\spa1.2\spa2.3\spa4.5}\, .
\eeq
We note that this matrix element has the same sign as BDK, Eq.~(D1), as do all of the results
in this section.
The $\spa i.j $ and $\spb i.j $ are the normal spinor products for massless vectors, 
such that $\spa i.j \spb j.i =s_{ij}$. For details of their definition see refs.~\cite{Dixon:1996wi,Ellis:2011cr}. 
\subsection{Fermion loop corrections to the tree level amplitude}
The one-loop colour decomposition is given by 
\begin{eqnarray}
\label{qqgDecomp}
   \A{5}^{1\rm -loop}(1_{q}, 2_g,  3_{\qb}) 
  &=&  2 e^2 \, g^3 \biggl\{ 
 \bigl( -Q^q  + v_{L,R}^e v_{L,R}^q  \, \prop{Z}(s_{45}) \bigr)   \nonumber \\
& \times & (T^{a_2})_{i_1}^{~\ib_3} \biggl[ N_c  A_{5;1}(1_q,2_g, 3_{\qb})  
                           + \frac{1}{N_c} A_{5;2}(1_q,3_{\qb};2_g) \biggl] \nonumber \\
&  + & 2 \sum_{f=t,b} \Af \, v_{L,R}^e \, \prop{Z}(s_{45}) 
        (T^{a_2})_{i_1}^{~\ib_3}  \, A_{5;3}^{f}(1_q,3_{\qb};2_g)   \biggr\} \, .
\end{eqnarray}
The results for the functions $A_{5;1}$ and $A_{5;2}$ are given in BDK~\cite{Bern:1997sc}.
Since they do not involve fermion loops we do not repeat them here.  
The function $A_{5;3}$ contains the terms where
a $Z$ couples to a loop of quarks via the axial coupling, as shown in Fig.~\ref{floop5}, where our conventions for the 
overall coupling factors are given in Table~\ref{Feynmanrules}.
If we consider all quarks except the top to be massless then
there is a net contribution only for the third generation, because of the opposite weak isospin
of the up- and down-type quarks

The result for the leading order interfered with the NLO and summed over colours is
given in terms of partial amplitudes as,
\begin{eqnarray}
 \sum_{{\rm colors}} [\A{5}^* \A{5}]_{{\rm NLO}} 
  & = & 8 e^4 \, g^4 \, (N_c^2-1)\,N_c \, {\rm Re} \Biggl\{ 
   \bigl( -Q^q  + v_{L,R}^e v_{L,R}^q \, \prop{Z}^*(s_{45}) \bigr) 
   A_5^{{\rm tree}*}(1_q,2_g,3_{\qb}) \nonumber \\
& \times &\biggl[   
  \bigl( -Q^q  + v_{L,R}^e v_{L,R}^q  \, \prop{Z}(s_{45}) \bigr)
  \bigl[A_{5;1}(1_q,2_g,3_{\qb}) + \frac{1}{N_c^2} A_{5;2}(1_q,3_{\qb}; 2_g) \bigr]\;  \nonumber \\
& + & \frac{2}{N_c} \sum_{f=t,b}  \Af \,v_{L,R}^e\, \prop{Z}(s_{45}) A_{5;3}^f(1_q,3_{\qb};2_g) \biggr] \Biggr\} \, .
\label{SquaredAmplitudeTwoQuarksOneGluon}
\end{eqnarray}
Note that the axial part $A_{5;3}^f$ depends on the flavour of the quark ($f$) and
we have to sum over the contributions of the top and bottom loops.
We choose to follow the conventions of the original BDK presentation and write the subleading contributions
with permuted momentum labels.  In this scheme we further define
\beq
A^f_{5;3}(1_q, 2_{\qb}; 3_g)
 = i \cg A_{\ax}^{f} (1_q^+, 2_{\qb}^-; 3_g^+)\, ,
\eeq
with 
\beq \label{cgdefn}
\cg = \frac{1}{(4 \pi)^{2-\epsilon}}\frac{\Gamma(1+\epsilon)\Gamma^2(1-\epsilon)}{\Gamma(1-2 \epsilon)}\, .
\eeq
In this case the terms of order $\epsilon$ and higher in $\cg$ are not needed because the amplitude is finite.
The result for this amplitude is, including both the top and bottom contributions,
\beq
A_{\ax}^{t,b} (1_q^+, 2_{\qb}^- ; 3_g^+)  = 2 \frac{\spb5.3 \spb3.1 \spa2.4 }{s_{45}} \Big[f(m_t; 0,s_{12},s_{45})-f(m_b; 0,s_{12},s_{45})\Big]\, ,
\eeq
where $m_f$ is the mass of the quark running in the triangular loop. We shall take the bottom quark to
be massless, $m_b=0$.

The function $f$ is the axial triangle function that depends on $m_f$, for which
results have been given in ref.~\cite{Bern:1997sc} and are detailed in Appendix~\ref{axialtriangle}.
For a massive quark, such as the top quark, in the special case where one of the legs of the triangle is light-like, 
we have, (c.f.~Eq.(\ref{Axial_function_massless}))
\begin{eqnarray}
f(m;0,q_1^2,q_3^2) &=&\frac{1}{2(q_3^2-q_1^2)}
\Bigg[1+2 m^2 C_0(q_1,q_3;m,m,m) \nonumber \\
 &+&\Big(\frac{q_3^2}{(q_3^2-q_1^2)}\Big)
 \Big[B_0(q_3;m,m)-B_0(q_1;m,m)\Big]\Bigg]\, .
\end{eqnarray}
$B_0$ and $C_0$ are the scalar integral functions defined in Appendix~\ref{app:scalarintegrals}.
In the limit $m \to \infty$ we get
\begin{equation}
f(m;0,q_1^2,q_3^2) = \frac{1}{24 m^2}\Big[1+\frac{(2 q_1^2+q_3^2)}{15 m^2}
          +\frac{(2 q_1^2 q_3^2+3 q_1^4+q_3^4)}{140 m^4}\Big] + O(1/m^8) \, .
\label{eq:f1jetlimit}
\end{equation}
The result for a massless quark is,
\begin{eqnarray}
f(0;0,q_1^2,q_3^2) &=&\frac{1}{2 (q_3^2-q_1^2)}
\Bigg[1+\frac{q_3^2}{(q_3^2-q_1^2)} \log\left(\frac{q_1^2}{q_3^2}\right)\Bigg] = \frac{1}{2 q_3^2} L_1\left(\frac{q_1^2}{q_3^2}\right) \, ,
\end{eqnarray}
where $L_0, L_1$ are the cut-completed functions,
\begin{equation}
\label{cutcompleted}
L_0(r) =\frac{\ln(r)}{1-r},\;\;L_1(r) =\frac{L_0(r)+1}{1-r}\, .
\end{equation}
Summing over the third generation isodoublet in the limit $m_b=0$ we get,
\beq
A_{\ax}^{t,b} (1_q^+, 2_{\qb}^- ; 3_g^+)  = 
\frac{\spb5.3 \spb3.1 \spa2.4 }{s_{45}}
 \Big[2 \, f(m_t; 0,s_{12},s_{45})-\frac{1}{s_{45}} L_1\left(\frac{-s_{12}}{-s_{45}}\right)\Big] \, .
\eeq
Keeping only the leading term in the $m \to \infty$ limit given in Eq.~(\ref{eq:f1jetlimit}), this agrees with
BDK, Eq.~(D.11).

\section{Six point amplitude, $A(1_q,2_{\Qb},3_{Q},4_{\qb},5_{\eb},6_e)$.}
\label{sec:sixpointqqQQ}

We now consider processes with one more parton in the final state, starting with processes containing
four quarks\footnote{We find that the overall sign for the six point processes, using the Feynman rules
of ref.~\cite{Ellis:1991qj}, is opposite to that of BDK. Since it is an overall sign it is of no importance;
to allow our results to be used as a supplement to BDK, we have adjusted our overall sign to agree with the conventions 
of BDK.}. 

\subsection{Tree graphs}
The general decomposition for the tree process requires that we include the two terms corresponding to 
the $Z/\gamma^*$ attaching to one or the other of the quark lines, 
\beqn
\A{6}^{\rm tree} (1_\q,2_{\Qb},3_Q,4_{\qb}) & = &
 2 e^2 g^2 \biggl[  \Bigl( - Q^\q 
+  v_{L,R}^e v_{L,R}^\q \,\prop{Z}(s_{56}) \Bigr) 
  \Atree_6(1_\q,2_{\Qb},3_Q,4_{\qb}) \nonumber \\
&+&  \Bigl( - Q^Q 
+ v_{L,R}^e v_{L,R}^Q \,\prop{Z}(s_{56})\Bigr) 
  \Atree_6(3_Q,4_{\qb},1_\q,2_{\Qb}) \biggr] \nonumber \\
& \times & 
\Bigl(\delta_{i_1}^{\ib_2} \,  \delta_{i_3}^{\ib_4}  \, 
-{1\over N_c} \delta_{i_1}^{\ib_4}\,  \delta_{i_3}^{\ib_2} \,\Bigr) \,.
\label{TreeQQQQColorDecomposition}
\eeqn
The result for the tree process is,
\beq
A^{\tree}_6(1_q^+,2_{\Qb}^+,3_Q^-,4_{\qb}^+,5_{\eb}^-,6_e^+)=\fudgeBDK i\Bigg[ 
  \frac{\spb1.2 \spa 4.5 \spab3.(1+2).6}{s_{23} s_{56} s_{123}}
 +\frac{\spa 3.4 \spb1.6 \spab5.(3+4).2}{s_{23} s_{56} s_{234}}\Bigg]\, .
\eeq
This result is in agreement with BDK Eq.~(12.3).

\subsection{One-loop results general structure}
The general structure of the decomposition at one loop is~\cite{Bern:1996ka}
\beqn
&&\A{6}^{1\rm -loop} (1_\q ,2_{\Qb},3_Q,4_{\qb})  =  \nonumber \\
&& 2 e^2 g^4  \biggl[\Bigl( - Q^\q 
+  v_{L,R}^e v_{L,R}^\q \prop{Z}(s_{56}) \Bigr) 
\Bigl[ N_c \,  \delta_{i_1}^{\ib_2} \, \delta_{i_3}^{\ib_4}  \, 
         A_{6;1}(1_\q,2_{\Qb},3_Q,4_{\qb})
   + \delta_{i_1}^{\ib_4}\,  \delta_{i_3}^{\ib_2} \,
           A_{6;2} (1_\q,2_{\Qb},3_Q,4_{\qb}) \Bigr] \nonumber \\
&+& \Bigl( - Q^Q 
+  v_{L,R}^e v_{L,R}^Q \,\prop{Z}(s_{56}) \Bigr) 
\Bigl[N_c \,  \delta_{i_1}^{\ib_2} \, \delta_{i_3}^{\ib_4}  \, 
         A_{6;1}(3_Q,4_{\qb},1_\q,2_{\Qb})
   + \delta_{i_1}^{\ib_4}\,  \delta_{i_3}^{\ib_2} \,
           A_{6;2} (3_Q,4_{\qb},1_\q,2_{\Qb}) \Bigr] \nonumber \\
&+& { v_{L,R}^e \over \sin2\theta_W }\prop{Z}(s_{56}) \,
\Bigl( \delta_{i_1}^{\ib_2} \, \delta_{i_3}^{\ib_4}  
     - {1\over N_c} \delta_{i_1}^{\ib_4}\,  \delta_{i_3}^{\ib_2} \Bigr)
      A_{6;3} (1_\q,2_{\Qb},3_Q,4_{\qb}) \biggr] \,.
\label{qqqqDecomp}
\eeqn
For the case of identical quark flavours ($q=Q$) see ref.~\cite{Bern:1996ka}.

We are only concerned with the terms containing heavy quark loops.
The formulas for the four-quark partial amplitudes,
$A_{6;i}(1_\q^+,2_{\Qb}^\pm,3_\Q^\mp,4_{\qb}^-)$,
expressed in terms of primitive amplitudes are
\beqn
 A_{6;1}(1_\q^+,2_{\Qb}^+,3_\Q^-,4_{\qb}^-) 
&=&  A^\nn_6(1,2,3,4)
- {2\over N_c^2} \bigl(A^\nn_6(1,2,3,4) +
                       A^\an_6(1,3,2,4) \bigr) 
+ {1\over N_c^2} A_6^{\sl}(2,3,1,4) \nonumber \\
&+& {n_s - n_{\!f} \over N_c} A^{s\,,\nn}_6(1,2,3,4)
- {n_{\! f} \over N_c} A^{\!f,\,\nn}_6(1,2,3,4) 
+ {1\over N_c} A_6^{t,\,\nn}(1,2,3,4) \,,\\
A_{6;2}(1_\q^+,2_{\Qb}^+,3_\Q^-,4_{\qb}^-) &=&  A^\an_6(1,3,2,4) 
+ {1\over N_c^2} \bigl(A^\an_6(1,3,2,4) +
                       A^\nn_6(1,2,3,4) \bigr)
- {1\over N_c^2} A_6^{\sl}(2,3,1,4) \nonumber \\
& - &{n_s - n_{\!f} \over N_c} A^{s,\,\nn}_6(1,2,3,4)
+ {n_{\! f} \over N_c} A^{\!f,\,\nn}_6(1,2,3,4) 
- {1\over N_c} A_6^{t,\,\nn}(1,2,3,4) \,,\\
A_{6;3}(1_\q^+,2_{\Qb}^+,3_\Q^-,4_{\qb}^-) &=& A^\ax_6(1,4,2,3) \,,
\label{nndecomp}
\eeqn
and
\beqn
 A_{6;1}(1_\q^+,2_{\Qb}^-,3_\Q^+,4_{\qb}^-) 
&=&  A_6^\an(1,2,3,4)
- {2 \over N_c^2}\bigl( A_6^\an(1,2,3,4) 
                      + A_6^\nn(1,3,2,4) \bigr)
- {1\over N_c^2}  A_6^{\sl}(3,2,1,4) \nonumber \\
& + &{n_s - n_{\!f} \over N_c} A^{s,\,\an}_6(1,2,3,4)
- {n_{\! f} \over N_c} A^{\!f,\,\an}_6(1,2,3,4) 
+ {1\over N_c} A_6^{t,\,\an}(1,2,3,4) \, , \\
A_{6;2}(1_\q^+,2_{\Qb}^-,3_\Q^+,4_{\qb}^-) 
&=&  A_6^\nn(1,3,2,4)
+ {1\over N_c^2} \bigl(  A^\nn_6(1,3,2,4) 
                       + A^\an_6(1,2,3,4) \bigr)
+ {1\over N_c^2}  A_6^{\sl}(3,2,1,4) \nonumber \\
& - &{n_s - n_{\!f} \over N_c} A^{s,\,\an}_6(1,2,3,4)
 + {n_{\! f} \over N_c} A^{\!f,\,\an}_6(1,2,3,4) 
- {1\over N_c} A_6^{t,\,\an}(1,2,3,4) \,,\nonumber \\
A_{6;3}(1_\q^+,2_{\Qb}^-,3_\Q^+,4_{\qb}^-) &=& -A^\ax_6(1,4,3,2)\, .
\label{andecomp}
\eeqn

\subsection{Top loops -- vacuum polarization contribution}
\label{sec:qqQQvacpol}

The one-loop contribution to the unrenormalized vacuum polarization is given by,
\beq
\Gamma^{\mu \nu}(p) = i g^2 \cg \Big[g^{\mu \nu} p^2-p^\mu p^\nu \Big] \pi(p^2) \, ,
\eeq
with $\cg$ given in Eq.~(\ref{cgdefn}). The contribution of a top quark loop 
to $\pi(p^2)$ is
\beq
\pi(p^2)=-\frac{4}{3} T_R \Big[ B_0(p,m,m)+\frac{2 m^2}{p^2} [B_0(p,m,m)-B_0(0,m,m)]-\frac{1}{3}\Big]\, ,
\eeq
where $T_R=\frac{1}{2}$.
Renormalization is effected by performing subtraction at zero momentum transfer ($p^2=0$), so that the effect of the 
top quark decouples at large momentum transfer.  
In this scheme both the running of the coupling and the evolution of the parton distributions
remain in the five flavour scheme.
We find
\beq
\pi(p^2)-\pi(0)=-\frac{2}{3} \Big[\big(1+ \frac{2 m^2}{p^2}\big) [B_0(p,m,m)-B_0(0,m,m)] +\frac{1}{3}\Big]\, .
\eeq

In this subtraction scheme, the renormalized contribution coming from the diagrams shown in Fig.~\ref{floopqqQQa} is,
\beqn
\label{qqQQTOPvacpolrenorm}
A^{t,+\pm}_6(1_q^+,2_{\Qb}^+,3_Q^-,4_{\qb}^+,5_{\eb}^-,6_e^+)&=&
 -\cg \frac{2}{3} \Big[ \left(1+\frac{2 m^2}{s_{23}}\right)
 (B_0(p_{23},m,m)-B_0(p_{3},m,m))+\frac{1}{3}\Big] \nonumber \\
&\times& A^{\tree}_6(1_q^+,2_{\Qb}^\pm,3_Q^\mp,4_{\qb}^+,5_{\eb}^-,6_e^+)\, .
\eeqn
Performing the large mass expansion, in the limit $m \to \infty$ we get,
\beqn
A^{t,+\pm}_6(1_q^+,2_{\Qb}^+,3_Q^-,4_{\qb}^+,5_{\eb}^-,6_e^+) &=&
 -\cg \Big[ \frac{2}{15} \frac{s_{23}}{m^2} 
+ \frac{1}{70} \Big(\frac{s_{23}}{m^2}\Big)^2 
+ \frac{2}{945} \Big(\frac{s_{23}}{m^2}\Big)^3+O\Big(\Big(\frac{s_{23}}{m^2}\Big)^4\Big) \Big] \nonumber \\
&\times& A^{\tree}_6(1_q^+,2_{\Qb}^\pm,3_Q^\mp,4_{\qb}^+,5_{\eb}^-,6_e^+)\, .
\eeqn
This result agrees with BDK, Eq.~(12.2).

\subsection{Top loops - axial vector coupling contribution}
\label{sec:qqQQaxial}

The contribution of the top and bottom quarks to the diagrams shown in Fig.~\ref{floopqqQQb} is,
\beqn
&&A_6^{\rm ax}(1^+_{q},2_{\Qb}^-,3_{Q},4_{\qb})= \fudgeBDK 
2 i \frac{1}{16 \pi^2}\frac{1}{s_{56}} \nonumber \\
&\times& \Big[
   \big(\frac{\spb6.3 \spa4.2 \spa2.5}{\spa1.2}-\frac{\spb6.1 \spb1.3 \spa4.5}{\spb1.2}\big) 
  (f(m_t;s_{12},s_{34},s_{56}) -f(m_b;s_{12},s_{34},s_{56})) \nonumber \\
 \nonumber \\
  &+&\big(\frac{\spb6.1 \spa2.4 \spa4.5}{\spa3.4}-\frac{\spb6.3 \spb3.1 \spa2.5}{\spb3.4}\big) 
 (f(m_t;s_{34},s_{12},s_{56})-f(m_b;s_{34},s_{12},s_{56}))\Big]\, .
\eeqn
The axial triangle function $f$ is presented in appendix~\ref{axialtriangle}.  In particular, the 
reduction of $f$ to scalar integrals for the case at hand is given in Eq.~(\ref{eq:fred}).

\section{Six point amplitude, $A(1_q,2_g,3_g,4_{\qb},5_{\eb},6_e)$.}
\label{sec:sixpointqqgg}

We now consider the process
\beq
0 \to q(p_1)+g(p_2)+g(p_3)+\bar{q}(p_4)+e^{+}(p_5)+e^{-}(p_6)\, , \\
\eeq
where we have adopted the labelling convention of BDK for this case.
The amplitudes for this process are most conveniently defined using the operation ${\rm exch}_{34}$,
which just represents the exchange of labels $3$ and $4$, 
\beq
{\rm exch}_{34}: \;\; 
3 \leftrightarrow 4
\label{exch34}
\eeq
as well as the following ``flip'' functions:
\beqn
&& {\rm flip}_1 : \; 1 \leftrightarrow 4,  \quad 2 \leftrightarrow 3, \quad 5 \leftrightarrow 6, \quad \spa i.j \leftrightarrow \spb i.j \,. 
\label{flip1} \\
&& {\rm flip}_2: \;\; 
1 \leftrightarrow 2,\;\; 
3 \leftrightarrow 4,\;\; 
5 \leftrightarrow 6,\;\;
\langle ij \rangle  \leftrightarrow [ij]
\label{flip2} \\
&& {\rm flip}_5 : \; 1 \leftrightarrow 2, \quad 5 \leftrightarrow 6, \quad \spa i.j \leftrightarrow \spb i.j \,. 
\label{flip5}
\eeqn
The latter symmetry operation is not defined in BDK, although it is a combination of ${\rm exch}_{34}$~(Eq.~(\ref{exch34}))
and ${\rm flip}_2$~(Eq.~(\ref{flip2})).
\noindent

\subsection{Tree graphs}
Following ref.~\cite{Bern:1997sc}, the colour decomposition of the tree-level contribution to $\A{6}$ is
\begin{equation}            
 \A{6}^{\tree} (1_{\q}, 2,3,4_{\qb} )         
  =  2 e^2 g^2 \bigl( -Q^q  + v_{L,R}^e v_{L,R}^q  \, \prop{Z}(s_{56}) \bigr)            
   \sum_{\sigma\in S_{2}}              
   (T^{a_{\sigma(2)}} T^{a_{\sigma(3)}})_{i_1}^{~{\ib}_4}\        
    \Atree_6 (1_{\q},\sigma(2),\sigma(3),4_{\qb})\,.                
\label{TreeColorDecomp}                   
\end{equation}
The independent results for helicities of the gluons in the tree amplitude are, c.f. BDK Eqs.~(8.4), (8.9)
and (8.15).
\beqn
&& -i A_6^{\rm tree} (1_q^+, 2^+_g , 3^+_g , 4_{\qb}^- )  =
   \fudgeBDK \frac{{\spa4.5}^2}{\spa1.2\spa2.3 \spa3.4\spa5.6} \, , \\
&& -i A_6^{\rm tree} (1_q^+, 2^+_g , 3^-_g , 4_{\qb}^-)  = \nonumber \\
&   \fudgeBDK& \biggl[
 {-\spa3.1\spb1.2 \spa4.5 \spab3.{(1+2)}.6 \over
   \spa1.2 s_{23} s_{123} s_{56}}
+ {\spa3.4\spb4.2 \spb1.6 \spab5.{(3+4)}.2 \over
    \spb3.4 s_{23} s_{234} s_{56}}
+ {\spab5.(3+4).2 \spab3.(1+2).6 \over
    \spa1.2 \spb3.4 s_{23} s_{56}} \biggr]\, , \\
&&-i A_6^{\rm tree} (1_q^+, 2^-_g , 3^+_g , 4_{\qb}^-)  = \nonumber \\
   &\fudgeBDK& \biggl[
 {\spb1.3^2\spa4.5 \spab2.{(1+3)}.6 \over
   \spb1.2 s_{23} s_{123} s_{56}}
- {\spa2.4^2\spb1.6 \spab5.{(2+4)}.3 \over
    \spa3.4 s_{23} s_{234} s_{56}}
- {\spb1.3\spa2.4\spb1.6\spa4.5 \over
    \spa3.4 \spb1.2 s_{23} s_{56}} \biggr]\, .
\eeqn
The remaining helicity combination may be obtained by combining the operations of parity (interchanging
$\spa i.j$ and $\spb i.j$) and charge conjugation (exchanging identities of external fermions
and anti-fermions).  Thus we have,
\beq
 A_6^{\rm tree} (1_q^+, 2^-_g , 3^-_g , 4_{\qb}^- ) =
  {\rm flip}_1 \left[ A_6^{\rm tree} (1_q^+, 2^+_g , 3^+_g , 4_{\qb}^- ) \right]\, ,
\eeq
where the operation ${\rm flip}_1$ is defined in Eq.~(\ref{flip1}) (and also BDK Eq.~(6.7)).

\subsection{General structure at one-loop}
The one-loop colour decomposition is given by~\cite{Bern:1997sc}
\begin{eqnarray}
\label{qqggDecomp}
& \A{6}^{1\rm -loop}(1_{q}, 2, 3,  4_{\qb}) 
  =  2 e^2 \, g^4 \biggl\{ 
\bigl( -Q^q  + v_{L,R}^e v_{L,R}^q  \, \prop{Z}(s_{56}) \bigr)  \nonumber \\
& \hskip 1. cm \times 
\biggl[ N_c\, \sum_{\sigma\in S_2} 
 (T^{a_{\sigma(2)}}T^{a_{\sigma(3)}})_{i_1}^{~\ib_4}
    \ A_{6;1}(1_q,\sigma(2),\sigma(3), 4_{\qb}) 
   + \delta^{a_2 a_3}
   \, \delta_{i_1}^{~\ib_4}
    \,  A_{6;3}(1_q,  4_{\qb}; 2, 3) \biggl] \nonumber \\ 
&   \hskip .5 cm 
+   \sum_{f=1}^{\nf} \Bigl( -Q^i + v_{L,R}^e 
             \Vf  \prop{Z}(s_{56}) \Bigr) \nonumber \\
& \hskip 2 cm  \times
   \Bigl[ (T^{a_2}T^{a_3})_{i_1}^{~\ib_4} 
   + (T^{a_3}T^{a_2})_{i_1}^{~\ib_4} 
   - {2\over N_c}\, \delta^{a_2 a_3}
       \, \delta_{i_1}^{~\ib_4} \Bigr] 
   \, A_{6;4}^\vect(1_q,4_{\qb}; 2, 3)  \nonumber \\
&   \hskip .5 cm 
+ \sum_{f=b,t} 2 \Af v_{L,R}^e\, \prop{Z}(s_{56}) \biggr[
  \sum_{\sigma\in S_2} 
  \Bigl( (T^{a_{\sigma(2)}} T^{a_{\sigma(3)}} )_{i_1}^{~\ib_4} 
   - {1\over N_c} \delta^{a_2 a_3} 
       \, \delta_{i_1}^{~\ib_4}  \Bigr)
   \, A_{6;4}^\ax(1_q, 4_{\qb}; \sigma(2), \sigma(3)) \nonumber \\
& \hskip 2 cm  
   + {1\over N_c} \delta^{a_2 a_3} \, \delta_{i_1}^{~\ib_4}
      \, A_{6;5}^\ax(1_q, 4_{\qb}; 2, 3) \biggr]
  \biggr\} \,,
\end{eqnarray}
where $Q^i$ is the electric charge (in units of the positron charge) of the $i$th
quark and $\nf$ is the number of light quark flavours.  
The partial amplitudes $A_{6;1}$ and $A_{6;3}$ represent contributions
where the $Z$ couples to the fermion line as shown in Fig.~\ref{floopa}.
The partial amplitudes $A_{6;4}^\vect$, $A_{6;4}^\ax$ and $A_{6;5}^\ax$
represent the contributions from a photon or $Z$ coupling to a
fermion loop through a vector or axial-vector coupling.
The full results with massless partons running in the loop have been given 
in BDK. The addition of this paper is to insert the full top quark mass dependence of
$A_{6;1}, A_{6;4}^\vect$, $A_{6;4}^\ax$ and $A_{6;5}^\ax$. 

The partial amplitudes were further decomposed in the original BDK paper into primitive
amplitudes as follows:
\begin{eqnarray}
  A_{6;1} (1_q, 2, 3, 4_{\qb}) &=& 
     A_6 (1_q, 2, 3, 4_{\qb})
  - {1\over N_c^2} A_6(1_q, 4_{\qb}, 3, 2)  \nonumber \\
 && 
  + {n_s -n_{\! f}\over N_c} A_6^s(1_q, 2, 3, 4_{\qb}) 
  - {n_{\! f} \over N_c} A_6^f (1_q, 2, 3, 4_{\qb}) 
  + {1\over N_c} A_6^t (1_q, 2, 3, 4_{\qb}) \,,\nonumber \\ 
 A_{6;3} (1_q, 4_{\qb}; 2, 3) & = & 
A_6(1_q, 2, 3, 4_{\qb}) + A_6(1_q, 3, 2, 4_{\qb}) +
A_6(1_q, 2, 4_{\qb}, 3) + A_6(1_q, 3, 4_{\qb}, 2) \nonumber \\
 &&
 +  A_6(1_q, 4_{\qb}, 2, 3) +  A_6(1_q, 4_{\qb}, 3, 2)\,,  \nonumber \\
 A_{6;4}^\vect (1_q, 4_{\qb}; 2,3) & =& -A_{6}^{\vect s} (1_q, 4_{\qb}, 2, 3)
         - A_{6}^{\vect f} (1_q, 4_{\qb}, 2, 3) \,, \nonumber \\
 A_{6;4}^\ax (1_q, 4_{\qb}; 2, 3) & = & A_{6}^\ax (1_q, 4_{\qb}, 2, 3) \,, \nonumber \\
 A_{6;5}^\ax (1_q, 4_{\qb}; 2, 3) & = & A_{6}^{\ax,\sl} (1_q, 4_{\qb}, 2, 3) \,.
\label{totalpartialamp}
\end{eqnarray}
We must therefore provide new expressions, containing the full top quark mass dependence,
for the following quantities:
\begin{itemize}
\item $A_6^t (1_q, 2, 3, 4_{\qb})$, in which the $Z$ boson couples to the light quark line.
\item $A_{6;4}^\vect (1_q, 4_{\qb}; 2,3)$, in which the $Z$ boson is radiated from a quark
loop through the vector coupling.  In our approach it is not useful to perform an additional
decomposition into $A_{6}^{\vect s}$ and $A_{6}^{\vect f}$.
\item $A_{6}^\ax (1_q, 4_{\qb}, 2, 3)$ and $A_{6}^{\ax,\sl} (1_q, 4_{\qb}, 2, 3)$, where
the $Z$ boson is radiated from a top or bottom quark loop through the axial coupling.
\end{itemize}
For the quantities $A_{6;4}^\vect$, $A_{6}^\ax$ and $A_{6}^{\ax,\sl}$ we will follow the conventions of the 
original BDK paper and not present expressions for the momentum labelling as in Eq.~(\ref{totalpartialamp}),
but instead do so for the configuration $(1_q, 2_{\qb}; 3, 4)$.

The colour sum for $e^+\,e^- \to \qb q gg$
in terms of partial amplitudes is,
\begin{eqnarray}
 \sum_{{\rm colors}} [\A{6}^* \A{6}]_{{\rm NLO}} 
  & = & 8 e^4 \, g^6 \, (N_c^2-1) {\rm Re} \Biggl\{ 
   \bigl( -Q^q  + v_{L,R}^e v_{L,R}^q \, \prop{Z}^*(s_{56}) \bigr) 
   A_6^{{\rm tree}*}(1_q,2,3,4_{\qb}) \nonumber \\
& \times &\biggl[   
  \bigl( -Q^q  + v_{L,R}^e v_{L,R}^q  \, \prop{Z}(s_{56}) \bigr)
  \bigl[ (N_c^2-1) A_{6;1}(1_q,2,3,4_{\qb}) \nonumber \\
& &
   - A_{6;1}(1_q,3,2,4_{\qb}) + A_{6;3}(1_q,4_{\qb}; 2,3) \bigr]\; 
   \nonumber \\
& + &  \sum_{f=1}^{\nf} \bigl( -Q^i + v_{L,R}^e \Vf \prop{Z}(s_{56}) \bigr)
   \Bigl(N_c-{4\over N_c} \Bigr) A_{6;4}^\vect(1_q,4_{\qb};2,3) \nonumber \\
& + &   \sum_{f=t,b}  2 v_{L,R}^e \Af \, \prop{Z}(s_{56})
  \Bigl[ \Bigl(N_c - {2\over N_c}\Bigr) A_{6;4}^\ax(1_q,4_{\qb}; 2,3) 
      -\frac{2}{N_c} A_{6;4}^\ax(1_q,4_{\qb}; 3,2) \nonumber \\
  & + & \frac{1}{N_c} A_{6;5}^\ax(1_q,4_{\qb}; 2,3) \Bigr] \biggr] \Biggr\} \; 
        + \;\{2 \leftrightarrow 3 \}
   \,. 
\label{SquaredAmplitudeTwoQuarksTwoGluons}
\end{eqnarray}

\subsection{Result for $\A{6}^{t} (1_{\q}, 2,3,4_{\qb})$}
\label{sec:qqggcouplelight}

The aim of this section is to calculate the full mass dependence of the quantity $A_6^{t}$,
which is part of $A_{6;1}$ that is defined in Eq.~(\ref{totalpartialamp}).  The relevant diagrams
do not contribute to $A_{6;3}$.  The minus sign for the fermion loop is included in $A_6^{t}$.
For this case the only non-zero amplitudes occur when the gluons have the same helicity.

The amplitude can be written as
\beq
\A{6}^{t} (1_{\q}, 2^+,3^+,4_{\qb} ) = \A{6}^{s} (1_{\q},2^+,3^+,4_{\qb} ) \times F^t(s_{23},m^2)\, ,
\eeq
where
\beqn
\A{6}^{s} (1_{\q},2^+,3^+,4_{\qb} ) &=& \mfudgeBDK i \frac{\cg}{3} \frac{1}{\spa2.3^2 s_{56}}
    \Big[-\frac{\spa4.5 \spba6.(1+2).3 \spb3.1}{s_{123}}
      +\frac{\spb1.6 \spab5.(4+2).3 \spa3.4 }{s_{234}}\Big]\, .
\eeqn
This agrees with BDK Eq.~(8.2).  The function $\A{6}^{s}$ is anti-symmetric under
the exchange of $2$ and $3$.
The mass-dependence enters through the function
\beqn
F^t(s_{23},m^2)&=& - \Big[1+6 m^2 C_0(p_2,p_3,m,m,m) 
+ \frac{12 m^2}{s_{23}}\Big(B_0(p_{23},m,m)-B_0(p_2,m,m)\Big)\Big]\, ,
\eeqn
which accounts for the effect of vertex and bubble corrections such as those shown in Fig.~\ref{floopa}.
In our renormalization scheme there is no net effect from top-quark self-energy corrections on external gluons.
The large mass expansion of $F^t(s,m^2)$ is
\beq
F^t(s,m^2)= \frac{1}{20} \frac{s}{m^2} + \frac{1}{210} \Big(\frac{s}{m^2}\Big)^2
+ \frac{1}{1680} \Big(\frac{s}{m^2}\Big)^3 + O\Big(\big(\frac{s}{m^2}\big)^4\Big)\, .
\eeq
After using this expansion the result for $A_6^t$ agrees with BDK Eq.~(8.3).

\subsection{Result for $\A{6;4}^{\vect} (1_{\q}, 2_{\qb}, 3, 4)$}
\label{sec:qqggvector}

The result for loops of massless quarks that couple via a vector coupling have been
given in BDK, in particular through their Eqs.~(11.1-11.2) and Eqs.~(11.5-11.7).  In
their approximation, which retains only terms of order $1/m_t^2$, the top quark
loop does not contribute since it enters only at order $1/m_t^4$ and beyond.
We therefore introduce the extra contribution of the top quark loop through,
\beq
\A{6;4}^{\vect} (1_{\q}, 2_{\qb}, 3, 4) = \A{6;4}^{\vect, BDK} (1_{\q}, 2_{\qb}, 3, 4)
 + \A{6;4}^{\vect,t} (1_{\q}, 2_{\qb}, 3, 4) \,.
\eeq

We will not present explicit results for the term $\A{6;4}^{\vect,t}$ since they
can be simply related to previously published results for the process
$gg \to ZZ$~\cite{Campbell:2013una}.  This exploits the fact that $\A{6;4}^{\vect,t}$
only receives contributions from box (not triangle) diagrams, so that replacing a single
$Z \to \ell \bar\ell$ current by a $g^* \to q\bar q$ one is trivial.  We have,
\beq
\A{6;4}^{\vect,t} (1_{\q}, 2_{\qb}, 3, 4)
 = - \left[ A_{LL}(3_g, 4_g, 1_e, 2_{\bar e}, 6_\mu, 5_{\bar \mu}) 
          + A_{LR}(3_g, 4_g, 1_e, 2_{\bar e}, 6_\mu, 5_{\bar \mu}) \right]\, .
\eeq
This is in accord with the procedure for extracting the vector-vector contribution
given in Eq.~(24) of Ref.~\cite{Campbell:2013una}, up to an expected change in the
overall factor and a sign to match the conventions of BDK.

\subsection{Result for $A^{\ax,\sl}(1_q,2_{\qb},3_g,4_g)$}
\label{sec:qqggaxialsl}

The sub-leading colour piece receives 
contributions from the diagram of the type shown in Fig.~\ref{floopb}a.
The full result for the third generation isodoublet is
\beq
-i A^{\ax,\sl}(1_q^+,2_{\qb}^-,3_g^+,4_g^+)=\mfudgeBDK \cg 
\frac{\spa2.5 \spb4.6 \spab2.(1+3).4}{\spa1.3 \spa2.3 s_{56}} 
\Big[2 f(m_t;0,s_{123},s_{56})-\frac{L_1(\frac{-s_{123}}{-s_{56}})}{s_{56}}\Big] + {\rm exch}_{34}\, ,
\eeq
where ${\rm exch}_{34}$ is defined in Eq.~(\ref{exch34}).
This expression agrees with BDK Eq.(11.4).
The result when the gluons have opposite helicities is,
\beq
-i A_6^{\ax,\sl}(1_q^+,2_{\qb}^-,3_g^+,4_g^-)=\mfudgeBDK \cg \frac{\spa2.4 \spa4.5 \spab2.(1+3).6}{\spa1.3 \spa2.3 s_{56}} 
\Big[2 f(m_t;0,s_{123},s_{56})-\frac{L_1(\frac{-s_{123}}{-s_{56}})}{s_{56}}\Big] + {\rm flip}_2 \; .
\eeq
The function $L_1$ is defined in Eq.~(\ref{cutcompleted}) and $f$ is defined in Eq.~(\ref{ffunctiondef}).
The swap flip$_2$ is defined in Eq.~(\ref{flip2}).%
\beq
A_6^{\ax,\sl}(1_q^+,2_{\qb}^-,3_g^-,4_g^+)=A_6^{\ax,\sl}(1_q^+,2_{\qb}^-,3_g^+,4_g^-)|_{3 \leftrightarrow 4}\, .
\eeq
This agrees with BDK Eq.(11.12).

\subsection{Result for $A^{\ax}(1_q,2_{\qb},3_g,4_g)$}
\label{sec:qqggaxial}

The most complicated case in which to account for the top-quark mass 
is the calculation of the leading-colour contribution from
a loop of massive fermions with an axial vector coupling to the $Z$-boson.
For a complete isodoublet of massless quarks there is no net contribution of this type since the
diagrams precisely cancel between the isospin partners.  For the $(t,b)$ isodoublet this
is no longer the case once a non-zero mass for the top quark is assumed.  The contribution of this
isodoublet has been presented, retaining only the leading $1/m_t^2$ terms in an expansion of the
top-quark diagrams, in the paper of BDK.  The result for the massless diagrams can be extracted from
their Eqs.~(11.3)-(11.4) and Eqs.~(11.8)-(11.12), simply by discarding the terms proportional to $1/m_t^2$.

Our base amplitude can be written as follows,
\beqn \label{Melrose}
-i A_6^{{\ax}} (1_q^+ , 2_{\qb}^- , 3_g^{h_3} , 4_g^{h_4} , 5_{\eb}^- , 6_e^+) &=&
 \sum_{x,y,z} d_{x|y|z}(3^{h_3} , 4^{h_4}) D_0^{x|y|z}
+\sum_{x,y} c_{x|y}(3^{h_3} , 4^{h_4}) C_0^{x|y} \nonumber \\ &&
+\sum_{x} b_{x}(3^{h_3} , 4^{h_4}) B_0^{x}
+R(3^{h_3} , 4^{h_4}) \,.
\eeqn
This is an expansion in terms of the scalar box ($D_0^{x|y|z}$), triangle ($C_0^{x|y}$) and bubble ($B_0^{x}$)
integrals, defined explicitly in Appendix~\ref{app:scalarintegrals}, as well as a left-over rational part ($R$).
The box and triangle coefficients in that expansion have a further mass expansion,
\beqn
d_{x|y|z}(3^{h_3} , 4^{h_4}) &=& d^{(0)}_{x|y|z}(3^{h_3} , 4^{h_4}) +m^2 d^{(2)}_{x|y|z}(3^{h_3} , 4^{h_4}) \, ,\\
c_{x|y}(3^{h_3} , 4^{h_4}) &=& c^{(0)}_{x|y}(3^{h_3} , 4^{h_4}) +m^2 c^{(2)}_{x|y}(3^{h_3} , 4^{h_4})\, ,
\label{massexp}
\eeqn
while the bubble coefficients and rational part are independent of the mass $m$.  We use this feature to simplify
the presentation of our results by replacing the expansion of Eq.~(\ref{Melrose}) by the more compact form,
\beqn \label{MelroseNew}
&& -i A_6^{{\ax}} (1_q^+ , 2_{\qb}^- , 3_g^{h_3} , 4_g^{h_4} , 5_{\eb}^- , 6_e^+) = 
 -i A_{6,BDK}^{{\ax}} (1_q^+ , 2_{\qb}^- , 3_g^{h_3} , 4_g^{h_4} , 5_{\eb}^- , 6_e^+)  \nonumber \\ && \qquad
+ d_{3|12|4}(3^{h_3} , 4^{h_4}) D_0^{3|12|4}
+ d_{4|3|12}(3^{h_3} , 4^{h_4}) D_0^{4|3|12}
+ d_{3|4|12}(3^{h_3} , 4^{h_4}) D_0^{3|4|12}  \nonumber \\ && \qquad
+ c_{3|4}(3^{h_3} , 4^{h_4}) C_0^{3|4}
+ c_{12|3}(3^{h_3} , 4^{h_4}) C_0^{12|3}
+ c_{12|4}(3^{h_3} , 4^{h_4}) C_0^{12|4}  \nonumber \\ && \qquad
+ c_{3|124}(3^{h_3} , 4^{h_4}) C_0^{3|124}
+ c_{4|123}(3^{h_3} , 4^{h_4}) C_0^{4|123}
+ c^{(2)}_{12|34}(3^{h_3} , 4^{h_4}) C_0^{12|34}\, .
\eeqn
The function $A_{6,BDK}^{{\ax}}$ collects the bubble and rational terms as well as the contribution from
the triangle coefficient $c^{(0)}_{12|34}$, all of which may be extracted from the previous calculation of BDK.
In the paper of BDK, the bubble coefficients have been 
re-organized to perform cut completion, leading to more compact expressions. It is thus more efficient to use this
compact form as our point of departure in presenting the results.  Note though that, in our case, the relevant completed
functions will be replaced by combinations of scalar bubble integrals that involve the internal top-quark mass. This is
a consequence of the fact that the bubble coefficients are unchanged in the massive case, but the integrals themselves 
are changed.

Apart from the contribution that can be extracted from the results of BDK, Eq.~(\ref{MelroseNew}) also enumerates all
of the remaining box and triangle integral coefficients that must be specified to complete the amplitudes.  Although
it appears that we should specify three box integral coefficients and six triangle coefficients this is not the case.
A number of relations between the various coefficients can be used to minimize the number of independent expressions
that must be given explicitly.  The simplest relations are those that just correspond to a relabelling of momenta,
for example,
\beq
d_{4|3|12}(3_g^{h_3} , 4_g^{h_4}) = - d_{3|4|12}(4_g^{h_4} , 3_g^{h_3}) \,.
\label{eq:34swap}
\eeq
An additional simplification is due to the structure of the infrared divergences that are present when $m=0$, which requires
that the box and triangle coefficients are related.  Explicitly, we make use of the identities,
\beq
c_{3|4}^{(0)}(3_g^{h_3} , 4_g^{h_4}) =
 -\frac{d_{4|3|12}^{(0)}(3_g^{h_3} , 4_g^{h_4})}{s_{123}}
 - \frac{d_{3|4|12}^{(0)}(3_g^{h_3} , 4_g^{h_4})}{s_{124}} \,,
\label{eq:IRc3_4}
\eeq
and
\beq
c_{123|4}^{(0)}(3_g^{h_3} , 4_g^{h_4}) = (s_{56}-s_{123}) \left[
 \frac{d_{4|3|12}^{(0)}(3_g^{h_3} , 4_g^{h_4})}{s_{34} s_{123}}
- \frac{d_{3|4|12}^{(0)}(3_g^{h_3} , 4_g^{h_4})}{s_{34} s_{124}}
-\frac{c_{12|4}^{(0)}(3_g^{h_3} , 4_g^{h_4})}{s_{124}-s_{12}} \right] \,,
\label{eq:IRc123_4}
\eeq
together with the partner relation that can be obtained by exchanging labels $3$ and $4$.
The coefficients of the $m^2$ term in the triangle expansion of Eq.~(\ref{massexp}), i.e. $c^{(2)}_{x|y}$, are related to the
rational part, $R$~\cite{Badger:2008cm}.\footnote{This relation would normally also involve the $m^4$ terms in the expansion of the box
integral coefficients, but they vanish in this case.}
We exploit this relation in order to determine the coefficient $c^{(2)}_{12|34}$ which would normally require much simplification
in an explicit analytic calculation,
\beq
c^{(2)}_{12|34}(3_g^{h_3} , 4_g^{h_4}) = 2 R(3_g^{h_3} , 4_g^{h_4})
 - c^{(2)}_{123|4}(3_g^{h_3} , 4_g^{h_4}) - c^{(2)}_{124|3}(3_g^{h_3} , 4_g^{h_4}) \,.
\label{eq:c12_34}
\eeq

Explicit results for the remaining independent coefficients will be given below.

\subsubsection{Box coefficients}

For the box coefficients $d^{(i)}$ it is sufficient to consider only
$3^+4^+$ and $3^+4^-$ helicity combinations.  The remaining helicities are
obtained from these ones according to,
\beq
d^{(i)}(3^-, 4^-) = -{\rm flip}_5 \left[ d^{(i)}(3^+, 4^+) \right] \,, \qquad
d^{(i)}(3^-, 4^+) = -{\rm flip}_5 \left[ d^{(i)}(3^+, 4^-) \right] \,,
\eeq
where ${\rm flip}_5$ is defined in Eq.~(\ref{flip5}).

\hfill \\
\paragraph{{$d_{3|12|4}$ coefficients:}}
\label{sec:d3_12_4}

The coefficients of the box integrals with gluons situated on opposite corners are:
\begin{eqnarray}
d^{(0)}_{3|12|4}(3^+,4^+)&=&
       \frac{(s_{123} s_{124}-s_{12} s_{56}) (\spa2.3 \spa4.5+\spa2.4 \spa3.5) \spa2.5}{4 \spa1.2 \spa3.4^3 \spa5.6} \\
d^{(2)}_{3|12|4}(3^+,4^+)&=& \frac{1}{2 s_{12} s_{56} \spa3.4^3} \Bigg[ \spa2.3^2 \spa4.5 \Big(
        \spab4.(1+2).4 \spb1.6 \spb2.3
       -\spab4.(2+3).6 \spb1.2 \spb3.4  \Big) \nonumber \\
     &+&\frac{1}{2} \spa2.3 \spab4.(1+2).3
        \Big( \spa3.4 \spa4.5 \spb1.4 \spb4.6
          + \spa3.4 \spa3.5 \spb1.4 \spb3.6 \nonumber \\
          &-& \spab4.(1+2).4 \spa3.5 \spb1.6
          - \spa3.4 \spa2.5 \spb1.2 \spb4.6
          - 2 \spa2.3 \spa4.5 \spb1.6 \spb2.4 \Big) \nonumber \\
     &+&\frac{1}{2} \spa2.3 \spa4.5 \spb1.6 \spab3.(1+2).4 \spab4.(1+2).3 \Bigg] - \Bigg[ 3 \leftrightarrow 4\Bigg]\, ,
\end{eqnarray}

\begin{eqnarray}
d^{(0)}_{3|12|4}(3^+,4^-)&=& 0 \, , \\
d^{(2)}_{3|12|4}(3^+,4^-)&=& \frac{1}{2 s_{12} s_{34} s_{56} \spb3.4} \Big[  
              -\spb1.3 \spb3.4^2 \spb3.6 \spa2.4 \spa3.4 \spa4.5 \nonumber \\
           &-& \frac{\spab4.(1+2).3}{2 \, \spab3.(1+2).4}   \Big(
               \spb1.3^2 \spb1.4 \spb4.6 \spa2.5 \spa1.3 \spa1.4
              + \spb2.3 \spb1.4 \spab5.(2+4).3 \spb4.6 \spa2.4 \spa2.3 \nonumber \\
              &-& \spb1.3 \spb1.4 \spb3.4 \spab2.(1+3).6 (\spa1.3 \spa4.5  - \spa1.5 \spa3.4)  \nonumber \\
              &-& \spb1.3 \spb1.4 \spb4.6 \spab1.(2+4).3 \spa2.3 \spa4.5
              + \spb2.3 \spb1.4 \spb3.6 \spab2.(1+3).4 \spa2.4 \spa3.5  \nonumber \\
              &+& 2 \spb1.4 \spb3.6 \spa4.5 \spab3.(2+4).3 \spab2.(1+3).4 
              - 2 \spb2.1 \spb3.4 \spb1.3 \spb4.6 \spa2.4 \spa2.5 \spa1.3 \Big) \nonumber \\
           &-& \frac{\spab4.(1+2).3}{2}   \Big(
               \spb1.3^2 \spb4.6 \spa2.4 \spa1.5
              - \spb1.3 \spb1.4 \spb3.6 \spa2.1 \spa4.5
              - 2 \spb1.3 \spb3.4 \spb3.6 \spa2.3 \spa4.5 \nonumber \\
              &+& \spab5.(2+3).1 \spb3.4 \spb3.6 \spa2.4
              + \spb1.3 \spb4.6 \spa2.4 \spab5.(2+4).3 \Big) \Big]\, .
\end{eqnarray}

\hfill \\
\paragraph{{$d_{3|4|12}$ coefficients:}}
\label{sec:d3_4_12}

The box integrals corresponding to two contiguous gluons have the following coefficients: 
\begin{eqnarray}
d^{(0)}_{3|4|12}(3^+,4^+)&=& 0 \, , \\
d^{(2)}_{3|4|12}(3^+,4^+)&=& 
        \frac{\spa2.5 \spb1.2 \spb3.4}{2\spa3.4 s_{12} s_{56}} \Bigg[
         \spa2.3 \spb3.6
       - \spab2.(1+4).6
        -\frac{s_{124}\spa2.3 \spb6.4}{\spab3.(1+2).4} 
       -\frac{s_{124}\spa2.4 \spb6.3}{\spab4.(1+2).3} \Bigg]\, ,
\end{eqnarray}

\begin{eqnarray}
d^{(0)}_{3|4|12}(3^+,4^-)&=&  \frac{1}{4} s_{34} s_{124} \Big[ 
      \frac{\spab3.(2+4).1^2 \spab5.(1+2).4^2-\spb1.4^2 \spa3.5^2 s_{124}^2}{\spb2.1 \spa6.5 \spab3.(1+2).4^4} \Big]\, , \\
d^{(2)}_{3|4|12}(3^+,4^-)&=&  
      \frac{\spa2.4 s_{124}}{2\spab3.(1+2).4 s_{12} s_{56}} \Big( 
        \spb2.1 \spb3.6 \spa2.5 + \spb1.3 \spb6.4 \spa4.5 \Big) \nonumber \\
       &+& 3 \frac{s_{124} \spa2.3 \spab4.(1+2).3 \spb4.6 \spab5.(2+4).1 }{2 \spab3.(1+2).4^2 s_{12} s_{56}} \nonumber \\
       &+& \frac{\spab4.(1+2).3}{2\spab3.(1+2).4 s_{12} s_{56}}  \Big(
            \spab5.(2+4).1 (\spb3.6 \spa2.3-\spb4.6 \spa2.4)
          - \spb1.6 \spa2.5 s_{124} \nonumber \\ && \qquad
          + \spb1.4 \spb3.6 \spa2.4 \spa3.5 \Big)
       - \frac{\spb1.3 \spb3.6 \spa2.4 \spa4.5}{2 s_{12} s_{56}}\, .
\end{eqnarray}

\subsubsection{Triangle coefficients}

In general there are six possible kinematic configurations of triangle integrals that may contribute
to this partial amplitude.  These are:
\begin{equation}
c_{3|4}, \quad
c_{12|3}, \quad
c_{12|4}, \quad
c_{12|34}, \quad
c_{4|123}, \quad
c_{3|124}, \quad
\end{equation}
where the third leg is clear from momentum conservation.  A summary of the method for determining
each of these coefficients is shown in Table~\ref{table:tricoeff}.  Note that, since the box integral coefficients
$d^{(0)}_{3|4|12}$ and $d^{(0)}_{4|3|12}$ vanish in the same-sign helicity amplitudes, the infrared relation
of Eq.~(\ref{eq:IRc3_4}) implies that $c^{(0)}_{3|4}(3^\pm, 4^\pm) = 0$.  The only coefficients that
remain to be given explicitly are $c^{(0)}_{12|3}$ and $c^{(2)}_{4|123}$, which will be specified in
Sections~\ref{sec:c12_3} and~\ref{sec:c123_4} respectively below. 
\begin{table}
\begin{tabular}{|l|l|l|}
\hline
 Coefficient & $c^{(0)}$ & $c^{(2)}$ \\
\hline
$c_{12|34}$ & extracted from BDK  & rational relation, Eq.~(\ref{eq:c12_34}) \\
$c_{12|3}$  & Section~\ref{sec:c12_3} & vanishes \\
$c_{12|4}$  & Section~\ref{sec:c12_3}+Eq.~(\ref{eq:34swap}) relabelling & vanishes \\
$c_{4|123}$ & infrared relation, Eq.~(\ref{eq:IRc123_4}) & Section~\ref{sec:c123_4} \\
$c_{3|124}$ & infrared relation, Eq.~(\ref{eq:IRc123_4}) & Section~\ref{sec:c123_4}+Eq.~(\ref{eq:34swap}) relabelling \\
$c_{3|4}$   & infrared relation, Eq.~(\ref{eq:IRc3_4})& vanishes \\
\hline
\end{tabular}
\caption{Determination of triangle coefficients. \label{table:tricoeff}}
\end{table}

\hfill \\
\paragraph{{$c_{12|3}$ coefficients:}}
\label{sec:c12_3}

For the triangle coefficients $c_{12|3}$ it is sufficient to consider only
$3^+4^+$ and $3^+4^-$ helicity combinations.  The remaining helicities are
obtained from these ones according to,
\beq
c_{12|3}(3^-, 4^-) = -{\rm flip}_5 \left[ c_{12|3}(3^+, 4^+) \right] \,, \qquad
c_{12|3}(3^-, 4^+) = -{\rm flip}_5 \left[ c_{12|3}(3^+, 4^-) \right] \,,
\eeq
where ${\rm flip}_5$ is defined in Eq.~(\ref{flip5}).
As indicated in Table~\ref{table:tricoeff}, the mass-dependent terms in the coefficient vanish:
\begin{equation}
c^{(2)}_{12|3}(3^+,4^+) = c^{(2)}_{12|3}(3^+,4^-) = 0\, .
\end{equation}
These triangle coefficients are thus fully-specified by,
\begin{eqnarray}
c^{(0)}_{12|3}(3^+,4^+)&=& -\frac{1}{2} \frac{(s_{123}-s_{12}) \spb1.2 \spb3.4}{{\spa3.4}^2 s_{12} s_{56}} \nonumber \\
&\times &\Big[\spa2.4^2 \spa3.5 \spb4.6
 +\spa2.3 \spa4.5 \spab2.(1+3).6
 +\spa2.4 \spa2.5 \spab3.(1+2).6\Big]\, , \\
c^{(0)}_{12|3}(3^+,4^-) &=& \Biggl(
   \spab3.(1+2).6 \spab5.(3+4).1  \spa1.2 \spb1.2 \Bigl(  - \spa1.2 \spb1.4
    + 2 \spa2.3 \spb3.4 \Bigr) \nonumber \\ &&
  - \spab3.(1+2).6  \spa2.3^2 
     \spa4.5 \spb1.2 \spb3.4^2 
  + 2 \spab5.(1+2).3  \spa2.3^2 \spb1.2 \spb4.6 \spab2.(1+3).2 
       \nonumber \\ &&
  + \spab5.(3+4).1  \spa2.3^2 \spb1.2 \Bigl( \spa1.2 \spb2.3 \spb4.6
   + \spa1.3 \spb3.4 \spb3.6 \Bigr)  \nonumber \\ &&
  + \spa1.2^3 \spa3.5 \spb1.2^2 \spb1.4 \spb1.6 + \spa1.2^2 \spa2.3 \spa2.5
   \spb1.2^3 \spb4.6 \nonumber \\ &&
  - 3 \spa1.2^2 \spa2.3 \spa3.5 \spb1.2^2 \spb1.6 \spb3.4
    + 3 \spa1.2 \spa2.3^2 \spa3.5 \spb1.2^2 \spb3.4 \spb3.6 \nonumber \\ &&
  - \spa1.2 \spa2.3^2
      \spa4.5 \spb1.2 \spb1.4 \spb2.3 \spb4.6 + \spa1.3  \spa1.5 \spa2.3^2
       \spb1.2 \spb1.3^2 \spb4.6 \nonumber \\ &&
  - \spa1.3 \spa2.3^2 \spa4.5 \spb1.2 \spb1.4 \spb3.4  \spb3.6
      - \spa2.3^3 \spa2.5 \spb1.2 \spb2.3^2 \spb4.6  \nonumber \\ && 
     + \spa2.3^3 \spa3.5 \spb1.2 \spb2.3 \spb3.4  \spb3.6
   \Biggr) \frac{s_{34} (s_{123}-s_{12})}{\spab3.(1+2).4^3}\, .
\end{eqnarray}

\hfill \\
\paragraph{{$c_{4|123}$ coefficients:}}
\label{sec:c123_4}

There are no simple symmetry relations between the $c_{4|123}$ coefficients of different helicities.  We must therefore
specify them all.

The coefficients that appear in the $m \to 0$ limit are simply obtained by using the infrared relation, Eq.~(\ref{eq:IRc123_4}).
The $c^{(2)}$ coefficients are more complicated:
\begin{eqnarray}
&& c_{4|123}^{(2)}(3_g^+ , 4_g^+) = \biggl[ \nonumber \\ && 
         2\frac{\spa3.4 \spb1.3 \spb4.6}{s_{123}} \Bigl(
            2 \spa1.2 \spa2.5 \spb1.2
          - \spa1.5 \spa2.4 \spb1.4
          - \spa2.4 \spa3.5 \spb3.4 
          + 2 \spa2.5 \spa3.4 \spb3.4
          \Bigr) \nonumber \\ &&
       - 4\frac{\spab2.(5+6).4 \spab4.(5+6).3}{\spab4.(5+6).4 s_{123}} \Bigl(
           \spa2.5 \spa3.4 \spb1.2 \spb4.6
          \Bigr) \nonumber \\ &&
       - 4\frac{\spab2.(5+6).4}{\spa1.3 \spab4.(5+6).4 s_{123}} \Bigl(
           \spa1.2 \spa1.5 \spa2.4 \spa3.4 \spb1.2^2 \spb4.6
          \Bigr)
       + \frac{\spab2.(5+6).4}{\spab3.(5+6).4} \spa3.4 \spb4.6 \Bigl(
            \spa4.5 \spb1.4
          - \spa2.5 \spb1.2
          \Bigr) \nonumber \\ &&
       + \frac{\spab2.(5+6).4}{\spab4.(5+6).3} \Bigl(
            \spa4.5^2 \spb1.3 \spb5.6
          \Bigr)
       + 4\frac{\spab3.(5+6).4}{\spa1.3 \spab4.(5+6).4 s_{123}}  \Bigl(
            \spa1.2 \spa2.4 \spa2.5 \spa3.4 \spb1.2 \spb2.3 \spb4.6
          \Bigr) \nonumber \\ &&
       + \frac{\spab4.(5+6).1}{\spab4.(5+6).3}   \Bigl(
            \spa1.2 \spa4.5 \spb1.3 \spb4.6
          \Bigr)
       + \frac{\spab4.(5+6).4}{\spab3.(5+6).4} \spa3.5 \spb1.4  \Bigl(
            2 \spa2.4 \spb4.6
          + \spa2.5 \spb5.6
          \Bigr) \nonumber \\ &&
       - 2\frac{\spab4.(5+6).4}{\spab4.(5+6).3}   \Bigl(
           \spa2.4 \spa4.5 \spb1.3 \spb4.6
          \Bigr) 
       - 4\frac{\spa1.2 \spa3.4 \spa5.6 \spb1.2 \spb4.6}{\spa1.3 \spab4.(5+6).4 s_{123}} \Bigl(
            \spa1.2 \spa2.4 \spb1.2 \spb4.6
          \Bigr) \nonumber \\ &&
       - 4\frac{\spa1.2 \spa2.5 \spa3.4 \spb1.2 \spb4.6 \spab4.(1+3).4}{\spa1.3 \spab4.(5+6).4}
       + 4\frac{\spa1.2 \spa2.5 \spa3.4 \spb1.2 \spb4.6}{\spa1.3}
       - 2\frac{\spa2.3 \spa4.5 \spb1.4 \spb4.6}{\spab3.(5+6).4} s_{123} \nonumber \\ && 
       - \frac{s_{123}}{\spab3.(5+6).4 \spab4.(5+6).4} \Bigl(
          \spa2.4 \spa3.4 \spa5.6 \spb1.4 \spb4.6^2
          \Bigr)
       + \frac{\spa2.4 \spa5.6 \spb4.6^2}{\spab3.(5+6).4} \Bigl(
            \spa3.4 \spb1.4
          - \spa2.3 \spb1.2
          \Bigr) \nonumber \\ &&
       - \frac{s_{123}}{\spab4.(5+6).3} \Bigl(
          \spa2.4 \spa4.5 \spb1.6 \spb3.4
          \Bigr)
       + \frac{s_{123}}{\spab4.(5+6).3 \spab4.(5+6).4} \Bigl(
            \spa2.4 \spa4.5^2 \spb1.4 \spb3.4 \spb5.6
          \Bigr) \nonumber \\ &&
       + \frac{\spa4.5 \spb5.6}{\spab4.(5+6).3} \Bigl(
            \spa1.2 \spa4.5 \spb1.3 \spb1.4
          + 2 \spa2.4 \spa2.5 \spb1.2 \spb3.4
          - \spa2.4 \spa3.5 \spb1.3 \spb3.4
          - \spa2.5 \spa3.4 \spb1.3 \spb3.4
          \Bigr) \nonumber \\ &&
       + 2\frac{\spa3.4 \spa5.6 \spb1.3 \spb4.6}{\spab4.(5+6).4 s_{123}} \Bigl(
            \spa1.3 \spa2.4 \spb1.3 \spb4.6
          - \spa1.2 \spa2.4 \spb1.4 \spb2.6 
          - \spa1.4 \spa2.4 \spb1.4 \spb4.6
          + \spa2.3 \spa2.4 \spb2.6 \spb3.4 \nonumber \\ && \quad
          - 2 \spa1.2 \spa3.4 \spb1.4 \spb3.6
          - 2 \spa1.4 \spa2.3 \spb1.3 \spb4.6
          + 2 \spa2.3 \spa3.4 \spb3.4 \spb3.6
          + \spa2.4 \spa3.4 \spb3.4 \spb4.6
          \Bigr) \nonumber \\ &&
       + 4\frac{\spa2.4 \spa4.5 \spb1.4 \spb4.6}{\spab4.(5+6).4} s_{123} 
       + \frac{\spb4.6}{\spab4.(5+6).4} \Bigl(
            4 \spa1.4 \spa2.5 \spa3.4 \spb1.3 \spb1.4
          + 2 \spa2.4^2 \spa5.6 \spb1.2 \spb4.6 \nonumber \\ && \quad
          - 2 \spa2.4 \spa2.5 \spa3.4 \spb1.2 \spb3.4
          + 2 \spa2.4 \spa2.5 \spa3.4 \spb1.4 \spb2.3
          \Bigr)
       + \spa2.4 \spa2.5 \spb1.2 \spb4.6
          - 2 \spa2.4 \spa4.5 \spb1.4 \spb4.6 \nonumber \\ &&
          - 3 \spa2.5 \spa4.5 \spb1.4 \spb5.6
          \biggr] / (4 \spa3.4^2 s_{12} s_{56}) \, ,
\end{eqnarray}
\begin{eqnarray}
&& c_{4|123}^{(2)}(3_g^- , 4_g^-) = \biggl[ \nonumber \\ && 
          2\frac{\spa4.5}{s_{123}} \Bigl(
            2 \spa1.2 \spa1.3 \spb1.3 \spb1.4 \spb1.6 
          + \spa1.2 \spa2.3 \spb1.3 \spb1.6 \spb2.4
          - 2 \spa1.2 \spa2.4 \spb1.2 \spb1.4 \spb4.6 \nonumber \\ && \quad
          - 2 \spa1.2 \spa3.4 \spb1.4^2 \spb3.6
          + 2 \spa1.3 \spa2.3 \spb1.3 \spb1.6 \spb3.4
          - \spa1.4 \spa2.3 \spb1.4^2 \spb3.6
          - \spa2.3^2 \spb1.2 \spb3.4 \spb3.6 \nonumber \\ && \quad
          + 2 \spa2.3^2 \spb1.3 \spb2.6 \spb3.4
          - 2 \spa2.3 \spa2.4 \spb1.4 \spb2.3 \spb4.6
          - \spa2.3 \spa2.4 \spb1.6 \spb2.4 \spb3.4
          + \spa2.3 \spa2.5 \spb1.2 \spb3.4 \spb5.6 \nonumber \\ && \quad
          + \spa2.3 \spa3.4 \spb1.4 \spb3.4 \spb3.6
          - 2 \spa2.3 \spa3.4 \spb1.6 \spb3.4^2
          + \spa2.3 \spa4.5 \spb1.4 \spb3.4 \spb5.6
          \Bigr) \nonumber \\ &&
       + \frac{\spab2.(5+6).4}{\spab3.(5+6).4} \spa4.5 \Bigl(
            2 \spa2.3 \spb1.2 \spb4.6
          - \spa1.3 \spb1.4 \spb1.6
          - \spa2.3 \spb1.4 \spb2.6
          - 4 \spa3.4 \spb1.4 \spb4.6
          \Bigr) \nonumber \\ &&
       - 2\frac{\spab4.(5+6).1 \spa1.2 \spa2.3 \spa4.5 \spb1.4 \spb2.6 \spb3.4}{\spab4.(5+6).4 s_{123}}
       + \frac{\spab4.(5+6).3}{s_{123}} \Bigl(
            2 \spa1.5 \spa2.3 \spb1.4 \spb1.6
          \Bigr) \nonumber \\ &&
       + 2\frac{\spab4.(5+6).3}{\spab4.(5+6).4 s_{123}} \spb1.4 \spb1.6 \Bigl(
            \spa1.3 \spa2.3 \spa4.5 \spb3.4
          + \spa1.4 \spa2.3 \spa5.6 \spb4.6
          - 2 \spa1.2 \spa1.3 \spa4.5 \spb1.4
          \Bigr) \nonumber \\ &&
       + 2\frac{\spab4.(5+6).4}{\spab4.(5+6).3} \spa4.5 \spb1.3  \Bigl(
            2 \spa2.4 \spb4.6
          + \spa2.5 \spb5.6
          \Bigr)
       + 4\frac{\spab5.(2+3).1}{\spab4.(5+6).4 s_{123}} \Bigl(
            \spa2.3 \spa3.4 \spa4.5 \spb3.4^2 \spb5.6
          \Bigr) \nonumber \\ &&
       - 8\frac{\spa1.2 \spa2.4 \spa4.5 \spb1.2 \spb1.4 \spb1.6 \spb3.4}{\spb1.3 \spab4.(5+6).4}
       + \frac{1}{\spab3.(5+6).4} s_{123} \spb1.4 \spb4.6 \Bigl(
            \spa2.3 \spa4.5 
          + 4 \spa2.5 \spa3.4 
          \Bigr) \nonumber \\ &&
       + \frac{\spa2.4 \spa3.4 \spa5.6 \spb1.4 \spb4.6^2}{\spab3.(5+6).4 \spab4.(5+6).4} s_{123}
       - \frac{\spa2.3 \spa5.6 \spb4.6}{\spab3.(5+6).4} \Bigl(
          \spa2.4 \spb1.2 \spb4.6
          + 4 \spa3.4 \spb1.4 \spb3.6
          \Bigr) \nonumber \\ &&
       + \frac{\spa2.4 \spa4.5 \spb1.4 \spb3.6 s_{123}}{\spab4.(5+6).3}
       - \frac{\spa2.4 \spa4.5^2 \spb1.4 \spb3.4 \spb5.6 s_{123}}{\spab4.(5+6).3 \spab4.(5+6).4}
       - 4\frac{\spa2.4 \spa4.5 \spb1.4 \spb4.6 s_{123}}{\spab4.(5+6).4} \nonumber \\ &&
       + 2\frac{\spa4.5 \spb3.4 \spb5.6}{\spab4.(5+6).3} \Bigl(
            2 \spa2.3 \spa4.5 \spb1.3
          + \spa2.4 \spa2.5 \spb1.2 
          + \spa2.5 \spa3.4 \spb1.3 
          \Bigr) \nonumber \\ &&
       + 2\frac{\spa4.5 \spb1.4 \spb5.6}{\spab4.(5+6).4 s_{123}} \Bigl(
            2 \spa1.2 \spa3.4 \spa4.5 \spb1.4 \spb3.4
          - \spa1.2 \spa2.3 \spa4.5 \spb1.2 \spb3.4
          - \spa1.4 \spa2.3 \spa5.6 \spb1.3 \spb4.6
          \Bigr)  \nonumber \\ && \quad
       + 2\frac{\spa4.5}{\spab4.(5+6).4} \Bigl(
            2 \spa2.3 \spa2.4 \spb1.2 \spb3.4 \spb4.6
          - 2 \spa2.5 \spa3.4 \spb1.4 \spb3.4 \spb5.6  \nonumber \\ && \quad
          - \spa1.2 \spa4.5 \spb1.4^2 \spb5.6
          - \spa1.4 \spa2.3 \spb1.3 \spb1.4 \spb4.6
          - \spa2.3 \spa4.5 \spb1.4 \spb3.4 \spb5.6
          \Bigr) \nonumber \\ &&
       + \spa2.3 \spa4.5 \spb1.3 \spb4.6
          + 4 \spa2.3 \spa4.5 \spb1.6 \spb3.4
          - \spa2.4 \spa2.5 \spb1.2 \spb4.6
          - 5 \spa2.4 \spa4.5 \spb1.4 \spb4.6  \nonumber \\ && \quad
          - \spa2.5 \spa3.4 \spb1.3 \spb4.6
          + \spa2.5 \spa4.5 \spb1.4 \spb5.6
          \biggr] / (4 \spb3.4^2 s_{12} s_{56})\, ,
\end{eqnarray}
\begin{eqnarray}
&& c_{4|123}^{(2)}(3_g^+ , 4_g^-)  = \biggl[ \nonumber \\ && 
          2\frac{\spa4.5}{s_{123}} \Bigl(
            3 \spa1.4 \spa2.4 \spb1.3 \spb1.6 \spb3.4
          - \spa1.2 \spa2.4 \spb1.3^2 \spb2.6
          - 2 \spa1.4 \spa2.5 \spb1.3^2 \spb5.6 \nonumber \\ && \quad
          - 4 \spa2.4^2 \spb1.2 \spb3.4 \spb3.6
          + 2 \spa2.4^2 \spb1.3 \spb2.6 \spb3.4
          - \spa2.4 \spa2.5 \spb1.3 \spb2.3 \spb5.6
          + \spa2.4 \spa3.4 \spb1.3 \spb3.4 \spb3.6
          \Bigr) \nonumber \\ &&
       - \frac{\spab2.(5+6).4 \spab4.(5+6).3 \spa2.5 \spa3.4 \spb1.2 \spb4.6}{\spab3.(5+6).4^2}
       + \frac{\spab2.(5+6).4 \spab4.(5+6).3 \spa4.5 \spb1.6}{\spab3.(5+6).4} \nonumber \\ &&
       + \frac{\spab2.(5+6).4}{\spab3.(5+6).4} \spa4.5 \spb3.6 \Bigl(
            \spa3.4 \spb1.3
          - \spa2.4 \spb1.2
          \Bigr)
       - \frac{\spab4.(5+6).1 \spab4.(5+6).3 \spa2.5 \spb4.6}{\spab3.(5+6).4} \nonumber \\ &&
       + 4\frac{\spab4.(5+6).1}{\spa1.3 \spab4.(5+6).4 s_{123}} \Bigl(
            \spa1.2 \spa1.4 \spa2.5 \spa4.5 \spb1.2 \spb3.4 \spb5.6
          \Bigr) \nonumber \\ &&
       - \frac{\spab4.(5+6).1}{\spab3.(5+6).4^2} s_{123}   \Bigl(
          \spa2.4 \spa3.5 \spb3.4 \spb4.6
          \Bigr) 
       + \frac{\spab4.(5+6).1}{\spab3.(5+6).4} \spa4.5 \spb3.4 \Bigl(
            3 \spa2.4 \spb4.6
          + \spa2.5 \spb5.6
          \Bigr) \nonumber \\ &&
       - 4\frac{\spab4.(5+6).3 \spab4.(5+6).4}{\spab3.(5+6).4^2}   \Bigl(
           \spa2.5 \spa3.4 \spb1.4 \spb4.6
          \Bigr)
       - \frac{\spab4.(5+6).3}{\spab3.(5+6).4}   \Bigl(
          \spa2.3 \spa4.5 \spb1.3 \spb4.6
          \Bigr) \nonumber \\ &&
       + 4\frac{\spab4.(5+6).3}{\spa1.3 \spab4.(5+6).4 s_{123}} \spa1.2 \spa3.4 \spa4.5 \spb1.2 \spb3.4 \Bigl(
            \spa1.2 \spb1.6
          - \spa2.3 \spb3.6
          \Bigr) \nonumber \\ &&
       + 3\frac{\spab4.(5+6).3}{\spab3.(5+6).4^2} s_{123}   \Bigl(
            \spa2.5 \spa3.4 \spb1.4 \spb4.6
          \Bigr)
       - \frac{\spab4.(5+6).3}{\spab3.(5+6).4^2} \spa2.4 \spa5.6 \spb4.6^2 \Bigl(
          \spa2.3 \spb1.2 
          + 4 \spa3.4 \spb1.4
          \Bigr) \nonumber \\ &&
       + 2\frac{\spab4.(5+6).3 \spa4.5}{\spab4.(5+6).4 s_{123}} \Bigl(
             \spa1.2 \spa2.4 \spb1.3 \spb1.4 \spb2.6
          + 2 \spa1.2 \spa3.4 \spb1.3 \spb1.6 \spb3.4
          + 2 \spa1.4 \spa2.5 \spb1.3 \spb1.4 \spb5.6 \nonumber \\ && \quad
          - \spa2.3 \spa2.4 \spb1.3 \spb2.6 \spb3.4
          - 2 \spa2.3 \spa3.4 \spb1.3 \spb3.4 \spb3.6
          - 2 \spa2.4 \spa2.5 \spb1.2 \spb3.4 \spb5.6
          + \spa2.4 \spa2.5 \spb1.3 \spb2.4 \spb5.6
          \Bigr) \nonumber \\ &&
       + \frac{\spab4.(5+6).4}{\spab3.(5+6).4} \spa4.5 \spb1.6 \Bigl(
            \spa2.4 \spb3.4
          - \spa1.2 \spb1.3
          \Bigr) 
       + 4\frac{\spa1.2 \spa2.4 \spa4.5 \spb1.2 \spb3.4}{\spa1.3 s_{123}} \Bigl(
            2 \spa1.4 \spb1.6
          + \spa3.4 \spb3.6
          \Bigr) \nonumber \\ &&
       + 4\frac{\spa1.2 \spa2.4 \spa4.5 \spb1.2 \spb3.4 \spb5.6}{\spa1.3 \spab4.(5+6).4 s_{123}} \Bigl(
            \spa2.5 \spa3.4 \spb2.3
          - \spa3.4 \spa4.5 \spb3.4
          - \spa1.4 \spa2.5 \spb1.2
          - 2 \spa1.4 \spa4.5 \spb1.4
          \Bigr) \nonumber \\ && \quad
       +\frac{\spa2.4 \spa3.4 \spa5.6 \spb1.3 \spb4.6^2 s_{123}}{\spab3.(5+6).4^2}
       + \frac{\spa2.4}{\spab3.(5+6).4} \Bigl(
            2 \spa2.4 \spa5.6 \spb1.2 \spb3.6 \spb4.6 \nonumber \\ && \quad
          - \spa2.5 \spa4.5 \spb1.2 \spb3.4 \spb5.6
          - \spa3.4 \spa5.6 \spb1.3 \spb3.6 \spb4.6
          - \spa4.5^2 \spb1.4 \spb3.4 \spb5.6
          \Bigr) \nonumber \\ &&
       + 2\frac{\spa2.4 \spa4.5 \spb3.4 \spb5.6}{\spab4.(5+6).4 s_{123}} \Bigl(
            4 \spa2.4 \spa4.5 \spb1.2 \spb3.4 
          - 3 \spa1.4 \spa4.5 \spb1.3 \spb1.4
          - 2 \spa1.4 \spa2.5 \spb1.2 \spb1.3 \nonumber \\ && \quad 
          - 2 \spa2.4 \spa2.5 \spb1.2 \spb2.3 
          - 2 \spa2.4 \spa4.5 \spb1.3 \spb2.4
          - \spa1.3 \spa4.5 \spb1.3^2
          - \spa3.4 \spa4.5 \spb1.3 \spb3.4
          \Bigr) \nonumber \\ &&
       - 2\frac{\spa2.4 \spa4.5^2 \spb1.3 \spb3.4 \spb5.6}{\spab4.(5+6).4}
       + \spa2.4 \spa4.5 \spb1.3 \spb3.6
          \biggr] / (4 \spa3.4 \spb3.4 s_{12} s_{56})\, ,
\end{eqnarray}
\begin{eqnarray}
&& c_{4|123}^{(2)}(3_g^- , 4_g^+)  = \biggl[ \nonumber \\ && 
         2\frac{\spa2.3 \spa3.4 \spb1.4 \spb4.6}{s_{123}} \Bigl(
            \spa1.5 \spb1.4
          - \spa3.5 \spb3.4
          \Bigr)
       - 4\frac{\spab2.(5+6).4}{\spab4.(5+6).4 s_{123}} \Bigl(
           \spa2.3 \spa3.4 \spa5.6 \spb1.2 \spb4.6^2
          \Bigr) \nonumber \\ &&
       + 4\frac{\spab3.(5+6).4}{\spab4.(5+6).3^2} \spa2.4 \spa4.5 \spb3.4 \spb5.6 \Bigl(
            \spa2.5 \spb1.2
          + \spa3.5 \spb1.3
          \Bigr) \nonumber \\ &&
       + \frac{\spab3.(5+6).4}{\spab4.(5+6).3} \Bigl(
            4 \spa2.3 \spa4.5 \spb1.3 \spb4.6
          + 2 \spa2.4 \spa2.5 \spb1.2 \spb4.6
          +   \spa2.4 \spa4.5 \spb1.4 \spb4.6 \nonumber \\ && \quad
          + 2 \spa2.5 \spa3.4 \spb1.3 \spb4.6
          + 3 \spa2.5 \spa4.5 \spb1.4 \spb5.6
          \Bigr) \nonumber \\ &&
       + 4\frac{\spab3.(5+6).4 \spa3.4 \spb4.6 }{\spab4.(5+6).4 s_{123}} \Bigl(
            \spa1.2 \spa2.5 \spb1.2 \spb1.4
          + \spa2.3 \spa2.5 \spb1.2 \spb3.4
          + \spa2.3 \spa3.5 \spb1.3 \spb3.4
          \Bigr) \nonumber \\ &&
       - 4\frac{\spab3.(5+6).4}{\spab4.(5+6).4}   \Bigl(
           \spa2.5 \spa3.4 \spb1.4 \spb4.6
          \Bigr)
       - 8\frac{\spa1.2 \spa2.5 \spa3.4 \spb1.2 \spb1.4^2 \spb4.6}{\spb1.3 \spab4.(5+6).4} \nonumber \\ &&
       - \frac{\spab4.(5+6).4}{\spab4.(5+6).3} \spa2.5 \Bigl(
            2 \spa2.3 \spb1.2 \spb4.6
          + 2 \spa3.4 \spb1.4 \spb4.6
          + \spa3.5 \spb1.4 \spb5.6
          \Bigr) \nonumber \\ &&
       + \frac{\spb1.4}{\spab4.(5+6).3} \Bigl(
            \spa1.2 \spa2.4 \spa3.5 \spb1.2 \spb4.6
          + \spa1.2 \spa3.4 \spa3.5 \spb1.3 \spb4.6
          - \spa2.4 \spa3.4 \spa5.6 \spb4.6^2  \nonumber \\ && \quad
          - 2 \spa1.2 \spa3.5 \spa4.5 \spb1.4 \spb5.6
          - 2 \spa2.3 \spa3.5 \spa4.5 \spb3.4 \spb5.6
          + \spa2.3 \spa4.5 \spa5.6 \spb4.6 \spb5.6
          \Bigr) \nonumber \\ &&
       + 2\frac{\spa3.4 \spb1.4 \spb4.6}{\spab4.(5+6).4 s_{123}} \Bigl(
            2 \spa1.2 \spa2.3 \spa3.5 \spb1.2 \spb3.4
          - 2 \spa1.2^2 \spa3.5 \spb1.2 \spb1.4 \nonumber \\ && \quad
          + 5 \spa1.2 \spa2.3 \spa5.6 \spb1.2 \spb4.6
          + \spa1.2 \spa2.3 \spa5.6 \spb1.4 \spb2.6
          + 2 \spa1.3 \spa2.3 \spa5.6 \spb1.4 \spb3.6 \nonumber \\ && \quad 
          + \spa1.4 \spa2.3 \spa5.6 \spb1.4 \spb4.6
          + \spa2.3^2 \spa5.6 \spb2.4 \spb3.6
          - \spa2.3 \spa3.4 \spa5.6 \spb3.4 \spb4.6
          \Bigr) \nonumber \\ &&
       + 2\frac{\spa3.4 \spb1.4 \spb4.6}{\spab4.(5+6).4} \Bigl(
            \spa2.3 \spa3.5 \spb3.4
          - \spa1.2 \spa3.5 \spb1.4
          - \spa1.3 \spa2.5 \spb1.4
          - \spa2.3 \spa5.6 \spb4.6
          \Bigr) \nonumber \\ &&
       - 3 \spa2.3 \spa3.5 \spb1.4 \spb4.6
          \biggr] / (4 \spa3.4 \spb3.4 s_{12} s_{56})\, .
\end{eqnarray}

\subsubsection{BDK contribution}

The final contribution in Eq.~(\ref{MelroseNew}) that must be specified is $A_{6,BDK}^{{\ax}}$.  This consists of terms
representing the contributions of the bubble integrals and rational terms, as well as the mass-independent coefficient of
the triangle $c_{12|34}$.  Although the bubble and triangle coefficients are the same as in the original BDK paper,
the integrals that they multiply are of course the ones with non-zero masses in the loop.   Our recasting therefore
necessitates the introduction of the following functions related to scalar bubble integrals,
\beqn
\Lt_{-1}(x,y,m^2)&=&  B_0^{p_y}-B_0^{p_x} \nonumber \\
\Lt_0(x,y,m^2)&=&  \frac{y}{(y-x)}\, \Lt_{-1}(x,y,m^2) \nonumber \\
\Lt_1(x,y,m^2)&=&  \frac{y}{(y-x)}\, \Big[\Lt_{0}(x,y,m^2)+1\Big]
\eeqn
such that $x=p_x^2,y=p_y^2$.  In the limit $m \to 0$ these reduce to the standard BDK functions,
\beq
\left. \Lt_i(x,y,m^2) \right|_{m \to 0} = L_i\left(\frac{-p_x^2}{-p_y^2}\right) \;,
\eeq
with $L_{-1}(x,y,0) \equiv \ln(-x)-\ln(-y)$.  The overall sign of our expressions is also opposite to the one of BDK, due to the fact
that our result describes the amplitudes for a top quark ($\tau_3^f=+1/2$) rather than the (massless) bottom quark
($\tau_3^f=-1/2$) in BDK.
As in the massless case, we only need consider three helicity combinations.  The
final one is related by,
\begin{equation}
A_{6,BDK}^{{\ax}} (1_q^+ , 2_{\qb}^- , 3_g^- , 4_g^- , 5_{\eb}^- , 6_e^+)
 = {\rm flip}_2 \left[ A_{6,BDK}^{{\ax}} (1_q^+ , 2_{\qb}^- , 3_g^+ , 4_g^+ , 5_{\eb}^- , 6_e^+) \right] \,.
\end{equation}

With the preliminaries  understood we can make use of BDK Eq.~(11.3) to write the amplitude with two gluons
of positive helicity as,
\begin{eqnarray}
 -i A_{6,BDK}^{{\ax}} (1_q^+ , 2_{\qb}^- , 3_g^+ , 4_g^+ , 5_{\eb}^- , 6_e^+) &=& \Big[
 -\frac{\spa2.5^2}{\spa1.2\spa5.6\spa3.4^2} {\Lt_{-1}}(s_{123},s_{56},m_t^2) \nonumber \\
&+&\frac{\spab2.4.6 \spa2.5 }{ \spa1.2 \spa3.4^2 s_{56}} 
\Bigg(\frac{s_{34}}{s_{56}} \Lt_1(s_{123},s_{56},m_t^2) 
+ \Lt_0(s_{123},s_{56},m_t^2) \Bigg) \nonumber \\
&+& \frac{\spab5.3.1 \spa2.5}{\spa5.6 \spa3.4^2} \frac{\Lt_{0}(s_{123},s_{12},m_t^2)}{s_{12}}-{\rm exch}_{34}\Big] \nonumber \\
&-&(s_{14}+s_{34})\frac{\spa2.5 \spb4.6}{\spa1.3 \spa3.4}
\frac{1}{s_{56}^2} \Lt_1(s_{123},s_{56},m_t^2) \nonumber \\
&-&\frac{\spab2.3.1 \spa2.5 \spb3.6 }{\spa2.4\spa3.4}
\frac{1}{s_{56}^2} \Lt_1(s_{124},s_{56},m_t^2)\, . 
\end{eqnarray}

The amplitudes with gluons of opposite helicity are not related by a symmetry, but do share a common
structure.  We note also that the recasting of BDK Eqs.~(11.9) and~(11.10) also requires the following
replacements to be made in the BDK formulae,
\begin{eqnarray}
&& Ls_{-1}^{2mh}(s_{34},s_{123},s_{12},s_{56}) \longrightarrow 
 \left( \frac{\delta_{34}}{2} + \frac{s_{12} s_{56}}{s_{123}} \right) I_3^{3m}(s_{12},s_{34},s_{56}) \, , \nonumber \\
&& I_3^{3m}(s_{12},s_{34},s_{56}) \longrightarrow -C_0^{12|34}\, .
\end{eqnarray}
in order to isolate the contribution of the triangle with three off-shell legs (c.f. BDK Eq.~(B.3)) in the notation
of this paper.  By adapting the formulae in this way we obtain,
\beqn
 -i A_{6,BDK}^{{\ax}} (1_q^+ , 2_{\qb}^- , 3_g^+ , 4_g^- , 5_{\eb}^- , 6_e^+) &=& -C^\ax
    +\frac{\spa2.4 \spa1.4 \spb4.6 \spab2.(1+3).6}{\spa1.2 \spa1.3 \spb5.6 \spab3.(1+2).4} 
    \frac{\Lt_1(s_{56},s_{123},m_t^2)}{s_{123}}\nonumber \\
    &+&\frac{\spab2.(1+3).6 \spab3.(1+2).6 \spb1.3}{\spb5.6 \spab3.(1+2).4^2} 
    \frac{\Lt_0(s_{123},s_{12},m_t^2)}{s_{12}} \nonumber \\
    &+&\frac{\spa2.4 \spab1.(2+3).4 \spab2.(1+3).6 \spab3.(1+2).6}{\spa1.2 \spa1.3 \spb5.6 \spab3.(1+2).4^2} 
    \frac{\Lt_0(s_{123},s_{56},m_t^2)}{s_{56}} \nonumber \\
    &-&\frac{\spa2.4 \spa3.5 \spab4.(1+3).6}{\spa1.3 \spa3.4 s_{56} \spab3.(1+2).4} +{\rm flip}_2 \, ,
\eeqn
and,
\beqn
 -i A_{6,BDK}^{{\ax}} (1_q^+ , 2_{\qb}^- , 3_g^- , 4_g^+ , 5_{\eb}^- , 6_e^+) &=& C^\ax (3\leftrightarrow 4)
    -\frac{\spb1.4^2 \spa4.5 \spab5.(2+3).1}{\spb1.2 \spb1.3 \spa5.6 \spab4.(1+2).3}
     \frac{\Lt_1(s_{56},s_{123},m_t^2)}{s_{123}} \nonumber \\ 
    &+&\frac{\spab5.(2+3).1 \spab5.(1+2).3 \spa2.3}{\spa5.6 \spab4.(1+2).3^2}
     \frac{\Lt_0(s_{123},s_{12},m_t^2)}{s_{12}} \nonumber \\ 
    &-& \frac{\spb1.4 \spab4.(2+3).1 \spab5.(2+3).1 \spab5.(1+2).3}{\spb1.2 \spb1.3 \spa5.6 \spab4.(1+2).3^2}
    \frac{\Lt_0(s_{123},s_{56},m_t^2)}{s_{56}}\nonumber \\
    &-&\frac{\spb1.4^2 \spa2.5 \spb3.6}{\spb1.3 \spb3.4 s_{56} \spab4.(1+2).3} +{\rm flip}_2\, .
\eeqn

The auxiliary common quantity is adapted from BDK Eq.~(11.9) and is given by,
\beqn
C^{\ax}&=& - \Big[-\frac{3}{2} \left(
            \spab5.2.1 \spab2.1.6
           +\spab5.6.1 \spab2.5.6  
           -\spab5.3.1  \spab2.4.6
           -\spab5.4.1 \spab2.3.6 \right)
            \frac{\spab4.(1+2).3}{\spab3.(1+2).4 \Delta_3} \nonumber \\ 
   &-&3 \frac{\delta_{34} (\spab5.2.1 \delta_{12}-\spab5.6.1 \delta_{56})\spab4.(1+2).3 \spab2.(1+3).6}{\spab3.(1+2).4 \Delta_3^2}
        -\frac{\spb1.3 \spa4.5 \spa2.4 \spb3.6}{\Delta_3} \nonumber \\ 
    &+& \frac{\spb1.4 \spa3.5 (s_{123}-s_{124}) \spab4.(1+2).3 \spab2.(1+3).6}{\spab3.(1+2).4^2 \Delta_3} 
     -\frac{1}{2} \frac{\spb1.3 \spa4.5 \spab2.(1+3).6}{s_{123} \spab3.(1+2).4} \nonumber \\    
    &-&\frac{1}{2} \frac{\spab2.(1+3).4^2 \spab3.(1+2).6^2-\spa2.3^2 \spb4.6^2 s_{123}^2}{\spa1.2 \spb5.6 \spab3.(1+2).4^4}
      \Big( \frac{\delta_{34}}{2} + \frac{s_{12} s_{56}}{s_{123}} \Big)  \Big]
      C_0^{12|34} \nonumber \\
  &+&C^{\ax}_1+C^{\ax}_1(1\leftrightarrow 6,2\leftrightarrow 5) \nonumber \\
  &+& \frac{\spab2.(1+3).6^2}{\spa1.2 \spb5.6 \spab3.(1+2).4^2} \Lt_{-1}(s_{56},s_{34},m_t^2) \\
  &+& \frac{\spa2.4 \spb3.6}{\spab3.(1+2).4} 
      \left(\frac{\spab2.4.6 \delta_{34}}{\spa1.2 \spb5.6 \Delta_3}
            -\frac{\spa2.4 \spa3.5 \delta_{56}}{\spa1.2 \spa3.4  \Delta_3}
   -\frac{\spb1.3 \spb4.6 \delta_{12}}{\spb3.4 \spb5.6  \Delta_3}
   -2 \frac{\spab5.3.1}{\Delta_3}  
    +\frac{\spa2.4 \spa3.5}{\spa1.2 \spa3.4 s_{56}}\right) \, , \nonumber
\eeqn
where the function $C^{\ax}_1$ is defined as,
\beqn
C^{\ax}_1 &=& \Bigg(
    -6 \frac{\spb1.2 \spab2.(1+3).6 (\spa2.5 \delta_{34}-2 \spa2.1 \spb1.6 \spa6.5) \spab4.(1+2).3}{\spab3.(1+2).4 \Delta_3^2}
    - \frac{\spb1.3 \spb4.6 \spab2.(1+3).6}{\spb3.4 \spb5.6 \spab3.(1+2).4^2} \nonumber \\
    &+&\spb1.4 
    \frac{\spab2.(1+3).6 (3 \spab3.(1+2).4 \spb3.6 -\spb4.6 (s_{123}-s_{124})) \spab4.(1+2).3}{\spb3.4 \spb5.6 \spab3.(1+2).4^2 \Delta_3}  \nonumber \\
    &-&\frac{\spb1.3 \spa2.4 \spb3.6^2}{\spb3.4 \spb5.6  \Delta_3}\Bigg) \Lt_{-1}(s_{12},s_{34},m_t^2)\, .
\eeqn
These functions are defined in terms of the additional quantities,
\begin{eqnarray}
&& \Delta_3=s_{12}^2+s_{34}^2+s_{56}^2-2 s_{12} s_{34}-2 s_{34} s_{56}-2 s_{56} s_{12}\, , \nonumber \\
&& \delta_{12}=s_{12}-s_{34}-s_{56},\; \delta_{34}=s_{34}-s_{56}-s_{12},\; \delta_{56}=s_{56}-s_{12}-s_{34} \,.
\end{eqnarray}

Finally, we note that the determination of the triangle coefficient $c^{(2)}_{12|34}$ using Eq.~(\ref{eq:c12_34})
requires knowledge of the rational part of the amplitudes.  We do not list these explicitly here since they
may be simply obtained from the expressions for $A_{6,BDK}^{{\ax}}$ through the relation,
\begin{equation}
 R(3^{h_3} , 4^{h_4}) = \left[ A_{6,BDK}^{{\ax}} (1_q^+ , 2_{\qb}^- , 3_g^{h_3} , 4_g^{h_4} , 5_{\eb}^- , 6_e^+)
  \right]_{C_0^{12|34} \to 0, \; \Lt_{-1}(x,y,m^2) \to 0} \nonumber \\
\end{equation}

\section{Definition of scalar integrals}
\label{app:scalarintegrals}

The scalar integrals themselves are defined as follows,
\beqn
&& B_0^x \equiv B_0(p_x;m_1,m_2)  
 =\frac{\mu^{4-d}}{i \pi^{\frac{d}{2}}\rG}\int d^d l \;
 \frac{1} {d(l,m_1) \, d(l+p_x,m_2)}
\nonumber \\
&& C_0^{x|y} \equiv C_0(p_x,p_y;m_1,m_2,m_3)  
= \frac{1}{i \pi^2}
\nonumber \\
&& \times \int d^4 l \;
 \frac{1} {d(l,m_1) \, d(l+p_x,m_2) \, d(l+p_x+p_y,m_3)} \\
&& D_0^{x|y|z} \equiv D_0(p_x,p_y,p_z;m_1,m_2,m_3,m_4) 
= \frac{1}{i \pi^2} \nonumber \\
&&
\times \int d^4 l \;
 \frac{1} {d(l,m_1) \, d(l+p_x,m_2) \, d(l+p_x+p_y,m_3) \, d(l+p_x+p_y+p_z,m_4)} 
\eeqn
where the denominator function is 
\beq
d(l,m)=(l^2-m^2+i\varepsilon) \, .
\eeq
For the purposes of this paper we take the masses in the
propagators to be real.  Near four dimensions we use $d=4-2 \epsilon$ (and for
clarity the small imaginary part which fixes the analytic
continuations is specified by $+i\,\varepsilon$).  
$\mu$ is a scale introduced so that the integrals
preserve their natural dimensions, despite excursions away from $d=4$.
We have removed the overall constant which occurs in $d$-dimensional integrals 
\beq
\rG\equiv\frac{\Gamma^2(1-\epsilon)\Gamma(1+\epsilon)}{\Gamma(1-2\epsilon)} = 
\frac{1}{\Gamma(1-\epsilon)} +{\cal O}(\epsilon^3) =
1-\epsilon \gamma+\epsilon^2\Big[\frac{\gamma^2}{2}-\frac{\pi^2}{12}\Big]
+{\cal O}(\epsilon^3)\,.
\eeq

\section{Numerical values of coefficients}

The test momenta are, in the notation $p = (E,p_x,p_y,p_z)$ (in GeV),
\begin{eqnarray}
       p_1&=&(-3.0,2.1213203435596424,1.0606601717798212,1.8371173070873839)\, , \nonumber \\
       p_2&=&(-3.0,-2.1213203435596424,-1.0606601717798212,-1.8371173070873839)\, ,\nonumber \\
       p_3&=&(0.85714285714285710,-0.31578947368421051,0.79685060448070799,0.0)\, ,\nonumber \\
       p_4&=&(2.0,2.0,0.0,0.0)\, ,\nonumber \\
       p_5&=&(1.0,-0.18421052631578949,0.46482951928041311,0.86602540378443860)\, ,\nonumber \\
       p_6&=&(2.1428571428571432,-1.5,-1.2616801237611210,-0.86602540378443860)\, .
\end{eqnarray}
with 
\begin{equation}
p_1+p_2+p_3+p_4+p_5+p_6=0 \,.
\end{equation}
We use $m_t=0.4255266775$~GeV.

The results for the various contributions to the  $A_6^{\ax}$ partial amplitudes are shown in
Tables~\ref{table:axslBDK}--\ref{table:axmmBDK}.  We show results for the non-zero box and triangle
coefficients as well as the remaining contribution $A_{6,BDK}^{\ax}$ that includes both
bubbles and rational terms.  The coefficient $c^{(0)}_{12|34}$ is not shown explicitly for the
amplitudes with opposite gluon helicity even though it is non-zero, since its effect is also included in
$A_{6,BDK}^{\ax}$.

\begin{table}
\begin{tabular}{|l|l|l|l|l|}
\hline
Coeff        &Re~$y^{(0)}(3^+,4^+)$ &Im~$y^{(0)}(3^+,4^+)$ &Re~$y^{(2)}(3^+,4^+)$ &Im~$y^{(2)}(3^+,4^+)$\\
\hline
$c_{3|124}$  &              &            & -0.13353418 & -0.49827218 \\
$c_{4|123}$  &              &            & -0.79126348 &\;0.38570625 \\
\hline
$b_{56}$     &\;0.20772009 &\;0.22131702 &             &             \\
$b_{123}$    & -0.13126100 &\;0.06398398 &             &             \\
$b_{124}$    & -0.07645909 & -0.28530101 &             &             \\
$R$          & -0.46239883 & -0.05628296 &             &             \\
\hline
\hline
Coeff&Re~$y^{(0)}(3^-,4^+)$&Im~$y^{(0)}(3^-,4^+)$&Re~$y^{(2)}(3^-,4^+)$&Im~$y^{(2)}(3^-,4^+)$\\
\hline
$c_{3|124}$  &             &              &\;0.20571266 & -0.0325607\\
$c_{4|123}$  &             &              & -0.29104980 &\;1.0411831\\
\hline
$b_{56}$     & -0.06950546& -0.15407601   &          &          \\
$b_{123}$    & -0.04828163& \;0.17271964  &          &          \\
$b_{124}$    & \;0.11778709& -0.01864363  &          &          \\
$R$          & -0.04266857& \;0.50431124  &          &          \\
\hline
\hline
 Coeff&Re~$y^{(0)}(3^+,4^-)$&Im~$y^{(0)}(3^+,4^-)$&Re~$y^{(2)}(3^+,4^-)$&Im~$y^{(2)}(3^+,4^-)$\\
\hline
$c_{3|124}$  &             &              &\;0.01389374 & -0.00477234\\
$c_{4|123}$  &             &              & -0.01145015 &\;0.07951003\\
\hline
$b_{56}$     & -0.00605585 & -0.01045720 &           &           \\
$b_{123}$    & -0.00189944 & \;0.01318975 &           &           \\
$b_{124}$    & \;0.00795529 & -0.00273255 &           &           \\
$R$          & \;0.00122180 & \;0.03736885 &           &           \\
\hline
\hline
Coeff&Re~$y^{(0)}(3^-,4^-)$&Im~$y^{(0)}(3^-,4^-)$&Re~$y^{(2)}(3^-,4^-)$&Im~$y^{(2)}(3^-,4^-)$\\
\hline
$c_{3|124}$  &              &             & \; 0.01389374  & -0.00477234\\
$c_{4|123}$  &              &             &   -0.01145015  &\;0.07951003\\
\hline
$b_{56}$     & \;0.03831823 & -0.00009208 &            &            \\
$b_{123}$    & -0.01341259 & \;0.00497689 &            &            \\
$b_{124}$    & -0.02490565 & -0.00488480 &             &            \\
$R$          & -0.06217526 & \;0.01073516 &            &            \\
\hline
\hline
\end{tabular}
\caption{Non-zero integral coefficients for the axial contribution to 
$A_6^{\rm ax,sl}(1_q^+,2_{\qb}^-,3_g,4_g)$. 
Only the contribution of the isospin $+\half$ quark is included.
\label{table:axslBDK}}
\end{table}

\begin{table}
\begin{tabular}{|r|r|r|r|r|}
\hline
 Coeff       &Re~$y^{(0)}(3^+,4^+)$& Im~$y^{(0)}(3^+,4^+)$ &Re~$y^{(2)}(3^+,4^+)$ &Im~$y^{(2)}(3^+,4^+)$\\
\hline
$d_{3|12|4}$ &    0.9847638139   &    -0.8139317869   &    -0.0826002899   &     0.2135941623   \\ 
$d_{4|3|12}$ &       &        &     0.0291379239   &    -0.5509736787   \\ 
$d_{3|4|12}$ &       &        &     0.2392276202   &     0.0407498609   \\ 
\hline
$c_{12|34}$  &       &        &     0.1143994146   &    -0.2174992557   \\ 
$c_{12|3}$   &    0.1299341143   &    -0.1073937774   &        &        \\ 
$c_{12|4}$   &    0.3031796001   &    -0.2505854807   &        &        \\ 
$c_{3|124}$  &   -0.0706659218   &     0.0584071421   &     0.0542191910   &     0.1995748531   \\ 
$c_{4|123}$  &   -0.2439114076   &     0.2015988454   &     0.1822387437   &    -0.0369792184   \\ 
\hline
$A^{\ax}_{6,BDK}$ & 0.0744415301   &    -0.0504750372   & & \\
\hline
\end{tabular}
\caption{Non-zero box and triangle coefficients and $A_{6,BDK}^{\ax}$ contribution for the
partial amplitude $A_6^{\ax}(1_q^+,2_{\qb}^-,3_g^+,4_g^+)$.
\label{table:axppBDK}}
\end{table}

 \begin{table}
 \begin{tabular}{|r|r|r|r|r|}
 \hline
 Coeff&Re~$y^{(0)}(3^-,4^+)$&Im~$y^{(0)}(3^-,4^+)$&Re~$y^{(2)}(3^-,4^+)$&Im~$y^{(2)}(3^-,4^+)$\\
 \hline
$d_{3|12|4}$ &      &        &    -0.6553781232   &     0.2267711354   \\ 
$d_{4|3|12}$ &   3.4035534642   &     4.4512143946   &     0.7044032221   &     0.0506388969   \\ 
$d_{3|4|12}$ &  -1.5958557084   &     0.0483030299   &    -0.5569345916   &     0.1188692057   \\ 
 \hline
$c_{12|34}$  &   --   &     --   &     0.0862624911   &     0.0311702697   \\ 
$c_{12|3}$   &   0.2901747505   &     0.3794945606   &        &        \\ 
$c_{12|4}$   &  -0.6802846449   &     0.0205907147   &        &        \\ 
$c_{3|124}$  &   0.1585625864   &    -0.0047993395   &    -0.0253157939   &     0.0094553225   \\ 
$c_{4|123}$  &  -0.5447140053   &    -0.7123845260   &     0.0537029872   &    -0.2691326008   \\ 
$c_{3|4}$    &   0.0006275632   &    -0.1771280345   &        &        \\ 
 \hline
$A^{\ax}_{6,BDK}$ & 0.2424976515   &     0.0640430134   & & \\
 \hline
 \end{tabular}
\caption{Non-zero box and triangle coefficients and $A_{6,BDK}^{\ax}$ contribution for the
partial amplitude $A_6^{\ax}(1_q^+,2_{\qb}^-,3_g^-,4_g^+)$.  Note that the coefficient $c^{(0)}_{12|34}$ is non-zero,
but not listed explicitly here since it is included in $A_{6,BDK}^{\ax}$.
\label{table:axmpBDK}}
 \end{table}

\begin{table}
\begin{tabular}{|l|l|l|l|l|}
\hline
 Coeff&Re~$y^{(0)}(3^+,4^-)$&Im~$y^{(0)}(3^+,4^-)$&Re~$y^{(2)}(3^+,4^-)$&Im~$y^{(2)}(3^+,4^-)$\\
\hline
$d_{3|12|4}$ &      &        &     0.0211964344   &     0.0565331937   \\ 
$d_{4|3|12}$ &  -1.6787391821   &     3.5273786346   &    -0.0813082954   &     0.1402683777   \\ 
$d_{3|4|12}$ &  -0.0292425126   &    -0.0031829012   &    -0.0630934240   &    -0.0104948556   \\ 
\hline
$c_{12|34}$  &  --   &     --   &    -0.0042375804   &     0.0022554921   \\ 
$c_{12|3}$   &  -0.1431232764   &     0.3007316400   &        &        \\ 
$c_{12|4}$   &  -0.0124655582   &    -0.0013568136   &        &        \\ 
$c_{3|124}$  &   0.0029055061   &     0.0003162498   &    -0.0025971585   &    -0.0016437585   \\ 
$c_{4|123}$  &   0.2686700101   &    -0.5645313242   &    -0.0144060528   &    -0.0236227320   \\ 
$c_{3|4}$    &   0.0677211776   &    -0.1369105940   &        &        \\ 
\hline
$A^{\ax}_{6,BDK}$ &  -0.0513599766   &     0.1115536722   & & \\
\hline
\end{tabular}
\caption{Non-zero box and triangle coefficients and $A_{6,BDK}^{\ax}$ contribution for the
partial amplitude $A_6^{\ax}(1_q^+,2_{\qb}^-,3_g^+,4_g^-)$.  Note that the coefficient $c^{(0)}_{12|34}$ is non-zero,
but not listed explicitly here since it is included in $A_{6,BDK}^{\ax}$.
\label{table:axpmBDK}}
\end{table}

 \begin{table}
 \begin{tabular}{|r|r|r|r|r|}
 \hline
 Coeff&Re~$y^{(0)}(3^-,4^-)$&Im~$y^{(0)}(3^-,4^-)$&Re~$y^{(2)}(3^-,4^-)$&Im~$y^{(2)}(3^-,4^-)$\\
 \hline
$d_{3|12|4}$ &    0.3650137298   &     1.8497925731   &     0.0537351845   &     0.2954518713   \\ 
$d_{4|3|12}$ &       &        &     0.1734167739   &     0.0931722390   \\ 
$d_{3|4|12}$ &       &        &    -0.0895807857   &     0.0584032726   \\ 
 \hline
$c_{12|34}$  &       &        &     0.0357377770   &     0.0646139200   \\ 
$c_{12|3}$   &    0.0481615338   &     0.2440698534   &        &        \\ 
$c_{12|4}$   &    0.1123769122   &     0.5694963246   &        &        \\ 
$c_{3|124}$  &   -0.0261931149   &    -0.1327397448   &     0.0196845554   &    -0.0097033679   \\ 
$c_{4|123}$  &   -0.0904084933   &    -0.4581662160   &    -0.0075863869   &    -0.0498510937   \\ 
$c_{3|4}$    &       &        &        &        \\ 
\hline
$A^{\ax}_{6,BDK}$ &  0.0071292120   &    -0.0070092524   & & \\
\hline
\end{tabular}
\caption{Non-zero box and triangle coefficients and $A_{6,BDK}^{\ax}$ contribution for the
partial amplitude $A_6^{\ax}(1_q^+,2_{\qb}^-,3_g^-,4_g^-)$.
\label{table:axmmBDK}}
\end{table}

\section{Axial triangle}
\label{axialtriangle}

\begin{figure}
\includegraphics[angle=270,width=0.8\textwidth]{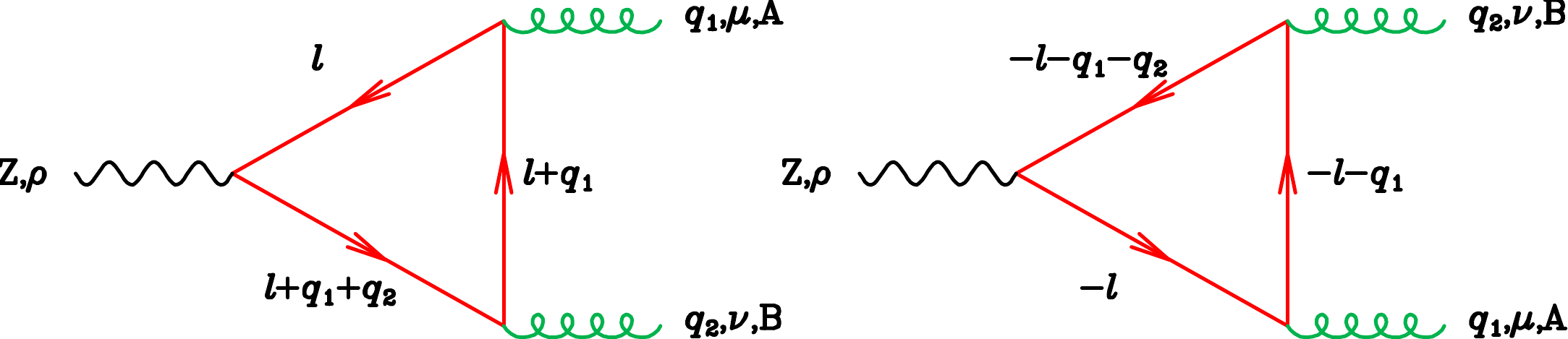}
\caption{Triangle graphs with an axial coupling to the $Z$-boson.}
\label{fig:anomaly}
\end{figure}

The amplitude for a $Z$ coupling to two gluons is denoted by $T^{\mu \nu \rho}_{AB}$. We
calculate the triangle shown in Fig.~\ref{fig:anomaly}, where all
momenta are outgoing $q_1+q_2+q_3=0$ and $q_i^2 \neq 0$.

The result for the two triangle diagrams shown in Fig.~\ref{fig:anomaly}, (including the
minus sign for a fermion loop) is,
\begin{equation} \label{EqT}
T^{\mu \nu \rho}_{AB}(q_1,q_2) = i \frac{g^2 e }{16 \pi^2} 
\delta_{AB} \; 2 v_A^f\;  \Gamma^{\mu \nu \rho}\,,
\end{equation}
where $v_A^f$ is given in Table~\ref{Feynmanrules} and
\begin{equation}
\Gamma^{\mu \nu \rho}(q_1,q_2,m) = \frac{1}{2}\frac{1}{i \pi^2} \int \; d^d l
\; \mbox{Tr}\Big\{ \gamma^\rho \gamma_5 \frac{1}{\slsh{l}-m}
\gamma^\mu \frac{1}{\slsh{l}+\slsh{q}_1-m}
\gamma^\nu \frac{1}{\slsh{l}+\slsh{q}_1+\slsh{q}_2-m} \Big\}\,.
\end{equation}
The most general form of $\Gamma$ consistent with QCD gauge invariance,
\begin{equation}
q_{1}^{\mu} \Gamma_{\mu \nu \rho}=
q_{2}^{\nu} \Gamma_{\mu \nu \rho} =0 \; ,
\end{equation}
can be written as,
\begin{eqnarray}
\Gamma^{\mu \nu \rho}&=& G_1 \;
      \Big\{ \mbox{Tr}[\gamma^\rho \gamma^\nu \slsh{q_1} \slsh{q_2}\gamma_5 ] q_1^\mu
            +\mbox{Tr}[\gamma^\rho \gamma^\mu \gamma^\nu \slsh{q_2}\gamma_5 ] q_1^2 \Big\}\nonumber \\
  &+& G_2 \;
        \Big\{\mbox{Tr}[\gamma^\rho\gamma^\mu\slsh{q_2}\slsh{q_1}\gamma_5 ] q_2^\nu
        +\mbox{Tr}[\gamma^\rho\gamma^\nu\gamma^\mu\slsh{q_1}\gamma_5 ] q_2^2 \Big\}\nonumber \\
  &+& G_3\; (q_1^\rho+q_2^\rho)
     \Big\{ \mbox{Tr}[ \gamma^\mu\gamma^\nu\slsh{q_1}\slsh{q_2}\gamma_5 ] \Big\}\nonumber \\
  &+& G_4\; (q_1^\rho-q_2^\rho)
      \Big\{ \mbox{Tr}[\gamma^\mu\gamma^\nu\slsh{q_1}\slsh{q_2}\gamma_5 ]  \Big\}\,. 
\end{eqnarray}
The functions $G_i$ are Lorentz invariant functions of $q_i^2,(i=1,3)$ and $m$.
By direct calculation it is found that $G_4=0$. To define the other $G_i$ we 
define the axial triangle function $f$
\beq
\label{ffunctiondef}
f(m;q_1^2,q_2^2,q_3^2)=\int_0^1 d^3a_i  \delta(1-a_1-a_2-a_3) 
\frac{a_2 a_3 }{[m^2-a_1 a_2 q_1^2-a_2 a_3 q_2^2-a_3 a_1 q_3^2]} \, .
\eeq
Full results for the function $f$ have been given in ref.~\cite{Bern:1997sc}.
We further define the integral 
\beq
I[j,k]=\int_0^1 d^3a_i  \delta(1-a_1-a_2-a_3) 
\frac{a_j a_k }{[m^2-a_1 a_2 q_1^2-a_2 a_3 q_2^2-a_3 a_1 q_3^2]}\, ,
\eeq
so that we have,
\beqn
G_1&=& f(m;q_2^2,q_1^2,q_3^2)=I[1,2] \nonumber \\
G_2&=& f(m;q_1^2,q_2^2,q_3^2)=I[2,3] \nonumber \\
G_3&=& f(m;q_1^2,q_3^2,q_2^2)=I[3,1]\; .
\eeqn
Contracting with the momentum of the
$Z$ boson we find that, 
\beq
(q_3)_\rho \, \Gamma^{\mu \nu \rho}= 
\Big[ -q_1^2 \, G_1 -q_2^2 \,G_2 -q_3^2 \, G_3\Big] 
  \mbox{Tr}[ \gamma^\mu\gamma^\nu\slsh{q_1}\slsh{q_2}\gamma_5 ]\,.
\eeq
The divergence of the axial current is easily seen to be,
\beq
(q_3)_\rho \, \Gamma^{\mu \nu \rho}=\Big[ m^2 C_0(q_1,q_2;m,m,m) +\frac{1}{2} \Big] 
\mbox{Tr}[ \gamma^\mu\gamma^\nu\slsh{q_1}\slsh{q_2}\gamma_5 ] \, ,
\eeq
showing the contribution of the pseudoscalar current proportional to $m^2$ and the anomalous term.
Summation over one complete quark doublet ($\tau_f = \pm 1/2$) cancels the anomaly term and solely the
piece proportional to the top-quark mass remains.

The function $f$ can be reduced to scalar integrals,
\beqn
f(m;q_1^2,q_2^2,q_3^2)&=&
      -\Big[3q_1^2 q_2^2 q_3^2 \frac{\delta_2}{\Delta_3^2}-\frac{(q_1^2 q_3^2-m^2 \delta_2)}{\Delta_3}\Big]
       C_0(q_1,q_2,m,m,m) \nonumber \\
      &+&\Big[3 q_1^2 q_3^2 \frac{\delta_3}{\Delta_3^2}-\frac{q_1^2}{2 \Delta_3} \Big]
       \Big(B_0(q_2,m,m)-B_0(q_1,m,m)\Big) \nonumber \\
      &+&\Big[(3 q_1^2 q_3^2 \frac{\delta_1}{\Delta_3^2}-\frac{q_3^2}{2 \Delta_3} \Big]
       \Big(B_0(q_2,m,m)-B_0(q_3,m,m)\Big) -\frac{1}{2} \frac{\delta_2}{\Delta_3}\, ,
\label{eq:fred}
\eeqn
in terms of the kinematic quantities,
\beqn
&&\delta_1=q_1^2-q_2^2-q_3^2 \,, \quad \delta_2=q_2^2-q_1^2-q_3^2 \,, \quad \delta_3=q_3^2-q_1^2-q_2^2 \,, \nonumber \\
&&\Delta_3=q_1^2 \delta_1+q_2^2 \delta_2+q_3^2 \delta_3 \,.
\eeqn
In the limit $q_1^2=0$ we get
\beqn
\delta_1=-q_2^2-q_3^2,\;\;\;\delta_2=-\delta_3=q_2^2-q_3^2,\;\;\;\Delta_3&=&(q_2^2-q_3^2)^2\, ,
\eeqn
and the result is,
\begin{eqnarray}
\label{Axial_function_massless}
f(m;0,q_2^2,q_3^2) &=&\frac{1}{2 (q_3^2-q_2^2)}
\Bigg[1+2 m^2 C_0(q_2,q_3;m,m,m) \nonumber \\
 &&+\frac{q_3^2}{(q_3^2-q_2^2)}
 \Big(B_0(q_3;m,m)-B_0(q_2;m,m)\Big)\Bigg]\,, \\
f(0;0,q_2^2,q_3^2) &=&\frac{1}{2 (q_3^2-q_2^2)}
\Bigg[1+\frac{q_2^2}{(q_3^2-q_2^2)} \log\left(\frac{q_2^2}{q_3^2}\right)\Bigg]\,.
\end{eqnarray}

When we are interested in the special case of an on-shell $Z$, with
$q_2^2=\varepsilon_2 \cdot q_2=0,\, \varepsilon_3 \cdot q_3=0$, then we only
get a contribution from $G_1$. 
The result for $G_1$ in this limit is $G_1=f(m;0,q_1^2,q_3^2)$. 

\end{document}